\documentclass[reprint,groupedaddress,showpacs,preprintnumbers,nofootinbib,nobibnotes,amsmath,amssymb,aps,prd,floatfix,superscriptaddress]{revtex4-1}

\usepackage{ifdraft}
\usepackage[colorlinks=false]{hyperref}
\usepackage{dcolumn}
\usepackage{bm}
\usepackage[normalem]{ulem}

\usepackage{siunitx}
\DeclareSIUnit\erg{erg}
\usepackage{pgfplots}
\usepgfplotslibrary[groupplots]
\usetikzlibrary{external}

\usepackage{graphicx,color,xcolor}
\usepackage{enumitem}
\usepackage[utf8]{inputenc}
\usepackage{mathrsfs}

\newcommand{\ar}[1]{#1}

\newcommand{\numberthis}{\addtocounter{equation}{1}\tag{\theequation}}

\def\mnras{Mon. Not. of the Royal Astron. Soc.} 	

\begin{document}

\title{Impact of electron capture rates on nuclei far from stability
  on core-collapse supernovae}

\author{Aur\'elien Pascal}\email{aurelien.pascal@obspm.fr}
\affiliation{LUTH, Observatoire de Paris, PSL Research University,
  CNRS, Universit\'e Paris Diderot, Sorbonne Paris Cit\'e, 5 place
  Jules Janssen, 92195 Meudon, France}%
\author{Simon Giraud}\email{giraud@ganil.fr}
\affiliation{Grand Acc\'el\'erateur National d'Ions Lourds (GANIL), CEA/DRF -
 CNRS/IN2P3, Boulevard Henri Becquerel, 14076 Caen, France}
\author{Anthea F. Fantina}\email{fantina@ganil.fr}
\affiliation{Grand Acc\'el\'erateur National d'Ions Lourds (GANIL), CEA/DRF -
 CNRS/IN2P3, Boulevard Henri Becquerel, 14076 Caen, France}
\author{Francesca Gulminelli}\email{gulminelli@lpccaen.in2p3.fr}
\affiliation{ LPC (CNRS/ENSICAEN/Universit\'e de Caen Normandie), UMR6534, 14050 Caen C\'edex, France}
\author{J\'er\^ome Novak}\email{jerome.novak@obspm.fr}
\affiliation{LUTH, Observatoire de Paris, PSL Research University,
  CNRS, Universit\'e Paris Diderot, Sorbonne Paris Cit\'e, 5 place
  Jules Janssen, 92195 Meudon, France}%
\author{Micaela Oertel}\email{micaela.oertel@obspm.fr}
\affiliation{LUTH, Observatoire de Paris, PSL Research University,
  CNRS, Universit\'e Paris Diderot, Sorbonne Paris Cit\'e, 5 place
  Jules Janssen, 92195 Meudon, France}%
\author{Adriana R. Raduta}\email{araduta@nipne.ro}
\affiliation{National Institute for Physics and Nuclear Engineering,  RO-077125 Bucharest, Romania}

\date{\today}

\begin{abstract}
  The impact of electron-capture (EC) cross sections on neutron-rich
  nuclei on the dynamics of core-collapse during infall and early
  post-bounce is studied performing spherically symmetric simulations
  in general relativity using a multigroup scheme for neutrino
  transport and full nuclear distributions in extended nuclear
  statistical equilibrium models. We thereby vary the prescription for
  EC rates on individual nuclei, the nuclear interaction for the EoS,
  the mass model for the nuclear statistical equilibrium distribution
  and the progenitor model. In agreement with previous works, we show
  that the individual EC rates are the most important source of
  uncertainty in the simulations, while the other inputs only
  marginally influence the results. A recently proposed analytic
  formula to extrapolate microscopic results on stable nuclei for EC
  rates to the \ar{high densities and temperatures and the }neutron
  rich region, with a functional form motivated by nuclear-structure
  data and parameters fitted from large scale shell model
  calculations, is shown to lead to a sizable (16\%) reduction of the
  electron fraction at bounce compared to more primitive prescriptions
  for the rates, leading to smaller inner core masses and slower shock
  propagation. We show that the EC process involves $\approx 130$
  different nuclear species around $^{86}$Kr mainly in the $N=50$
  shell closure region, and establish a list of the most important
  nuclei to be studied in order to constrain the global rates.
\end{abstract}


\maketitle

\section{Introduction}\label{s:intro}

Much effort has been devoted during decades to numerical simulations
of core-collapse supernovae (CCSN) and a lot of progress has been
achieved understanding the complex physics of these spectacular events
(see e.g.~\cite{Janka2012a, Janka2017b}).
But even if the main lines have been elucidated,
many details still deserve attention.

\ar{It has been first pointed out by \textcite{Bethe_NPA_1979} and
  confirmed by many subsequent studies that electron capture (EC) on
  nuclei plays an important role during the late stages of stellar
  evolution and the pre-bounce phase of CCSN~\cite{Aufderheide94,
    Heger2001a, Heger2001b, Hix2003, Janka2007}.}  For a very close
CCSN, DUNE~\cite{DUNE2018} might even be able to detect neutrinos from
the pre-bounce phase, as indicator of EC reactions~\cite{Kato2017}. \ar{Most sophisticated simulations of CCSN evolution thereby model EC in inhomogeneous nuclear matter typically by considering a nuclear statistical ensemble (NSE) distribution of nuclei together with microscopically calculated EC rates
  \cite{Juodagalvis_NPA_2010}. }

Different recent works have, however, pointed out that our
understanding of these processes under relevant thermodynamic
conditions is still insufficient and has an impact on the dynamics of
core collapse~\cite{Sullivan16, Raduta2016,
  Furusawa2017, Raduta2017, Yudin2018, Nagakura2019}. In particular,
the systematic study by Sullivan et al.~\cite{Sullivan16} has shown
that the uncertainties on the EC rates on individual nuclei induce
stronger modifications on the mass of the inner core at bounce and the
maximum of the neutrino luminosity peak than the progenitor model or
the equation of state (EoS).

The results of Ref.~\cite{Sullivan16} and the subsequent
study~\cite{Titus2017} clearly indicate that the simulations are most
sensitive to the EC rates for neutron-rich nuclei near the $N$ = 50
closed shell and to less extent to the next close shell at $N$ =
82. The main difficulty is that for the relevant nuclei not much
information is available, neither experimental nor from microscopic
calculations. \ar{The situation is nevertheless expected to improve in the near future due
to dedicated research programs (e.g.~\cite{Titus2019}). }Several other works have highlighted other
aspects. First of all, total EC rates are influenced as well by the
nuclear distribution given by the EoS as by the rates on individual
nuclei, \ar{suffering both from uncertainties.  Uncertainties in
  matter composition mainly stem from the definition of clusters in a
  hot nuclear environment and nuclear properties far from the
  stability valley.  Uncertainties in rates on individual nuclei are
  mainly due to nuclear structure and finite temperature effects. The
  sensitivity of EC rates to the former was addressed in
  Refs.~\cite{Raduta2016,Furusawa2017,Yudin2018}.  Thereby, in
  \textcite{Raduta2016} it has been shown that }the unknown binding
energies of nuclei far beyond the stability valley and a possible
shell quenching might increase the total EC rate by up to
30\%. Nuclear abundances also influence the neutrino opacity via
neutrino-nucleus scattering~\cite{Furusawa2017}. Future experiments
with exotic beams might improve the situation in that
respect, but further work is
necessary.\ar{\textcite{Nagakura2019} point out that a consistent
  treatment of nuclear abundances in the EoS and for calculating weak
  interaction rates is important to correctly study the EoS dependence
  of both the dynamics and neutrino signals.}

Concerning the individual rates, mostly values from microscopic
calculations -- where available~\cite{FFN_1982b, LMP_ADNDT_2001,
  Langanke2002, Oda1994, Pruet2003,Juodagalvis_NPA_2010}-- have been
used. \ar{By far the richest collection of microscopically calculated
  EC rates is thereby discussed in \textcite{Juodagalvis_NPA_2010}. In
  particular, in order to be able to extend the calculations to
  heavier and more neutron rich nuclei populated abundantly in the later
  stages of collapse, the authors define a strategy to describe
  electron captures by a hierarchy of nuclear models. However,
  although this seminal work of \textcite{Juodagalvis_NPA_2010} is used
  by some groups performing CCSN simulations, the data are not
  publicly available. The analytic parameterization of
  Ref.~\cite{Langanke_PRL_2003} is designed to complement microscopic
  data for high electron densities and temperatures and is extensively used in
  simulations, also to calculate rates for nuclei not present in the
  data bases.}

It has to be emphasized that the nuclei identified in
Refs.~\cite{Sullivan16,Titus2017} as having the highest impact lie
outside the region where \ar{--apart from the work by \textcite{Juodagalvis_NPA_2010}--} microscopic calculations exist. Therefore,
awaiting for more microscopic calculations and additional
information from charge-exchange experiments which e.g. should
correctly include additional Pauli blocking effects~\cite{Titus2017},
in Ref.~\cite{Raduta2017} an extended analytic parameterization has
been proposed incorporating different physical effects at high
electron density, temperature and isospin ratio with the aim of
improving the reliability of the extrapolation to regions not covered
by microscopic calculations. It has been shown that the improved
parameterization leads to a systematic reduction of the total EC rate
in agreement with expectations~\cite{Titus2017} which can reach one
order magnitude for some thermodynamic conditions.

In Refs.~\cite{Raduta2016,Raduta2017}, the impact of a potential
shell-quenching and of the improved EC rate parameterization on the
pre-bounce evolution of core collapse has been illustrated using some
typical thermodynamic conditions with EC rates added
perturbatively. Here we will perform self-consistent core-collapse
simulations investigating the effect of modified EC rates and the mass
model on the evolution. In contrast to Ref.~\cite{Sullivan16}, where
individual EC rates were thereby globally scaled by arbitrary factors
ranging from 2 to 10 with respect to the fiducial values, we show here
the effect of a physically motivated reduction of EC on nuclei. For an
easier use of our improved model in simulations, we will provide total
EC rates and neutrino-nucleus scattering opacities for the employed
EoS within the \textsc{Compose} data
base\footnote{\url{https://compose.obspm.fr}}~\cite{Typel2013},
\ar{with a numerically efficient format easily adaptable to any EoS}.

The paper is organized as follows. We start with specifying the setup
of our simulations in Section \ref{s:models}. In section
\ref{s:results} we discuss the influence of different ingredients on
the infall and early post-bounce evolution. In addition to comparing
the different prescriptions for the EC rates on individual nuclei, we
investigate the dependence on the progenitor model and on the EoS,
including different nuclear interaction models and different mass
models for determining binding energies of neutron rich
nuclei. Section \ref{s:relevant_nuclei} is devoted to a determination
of the most relevant nuclei for EC in order to specify the nuclei for
which microscopic and/or experimental studies are the most needed. We
conclude in Section \ref{s:conc}.

\section{Setup of the simulations}\label{s:models}

\subsection{General description}\label{ss:coconut}

In order to perform self-consistent numerical simulations of CCSN, we
use two different hydrodynamic codes in general relativity. Most
results are obtained with the spherically symmetric version of the
\textsc{CoCoNuT} code~\cite{coconut}. It solves the
general-relativistic hydrodynamics, with a 3+1 decomposition of
spacetime. High-resolution shock-capturing schemes are used for
hydrodynamic equations, whereas Einstein equation for gravitational
field is solved with spectral methods~\cite{dimmelmeier-05}. In
addition to the 5 evolution equations solved for hydrodynamics (coming
from the conservation of baryon current and energy-momentum tensor),
this model considers the advection equation for the electron fraction
$Y_e=n_e/n_B$, where $n_e$ and $n_B$ are the electron and baryon
number densities, respectively.

The sources terms for neutrino energy losses and deleptonization are
computed using the ``Fast Multigroup Transport'' scheme developed
by~\citet{mueller-15}. This scheme solves the stationary neutrino
transport in the ray-by-ray approximation, with a closure relation for
the first Eddington factor. In the collapse phase the first Eddington
factor is set to 1, which is equivalent to a free-streaming
condition. In the post-bounce phase the closure in obtained from a
two-stream approximation as in \cite{mueller-15}. We switch between
these two closures relations at neutrinosphere detection (when the
central optical depth becomes larger than 0.66).

The analyses of the mass model and the most relevant nuclei in
sections \ref{ss:mass_model} and \ref{s:relevant_nuclei} require the
use of a very flexible input for the EoS and the matter
composition. This is most easily achieved employing the perturbative
approach of Ref.~\cite{grams2018}. For simplicity, this has not been
implemented into the \textsc{CoCoNuT} code which works with tabulated
versions of the EoS, but in an improved version of the code developed
in Refs.~\cite{Romero1996,romero1996b,fantinaphd}, called
\textsc{ACCEPT} in the following. This spherically symmetric code uses
the same techniques for solving GR hydrodynamics as
\textsc{CoCoNuT}. The differences between both codes in solving
Einstein equations are negligible within our CCSN pre-bounce
context. Neutrinos are treated in a simple leakage-type scheme with a
multi-group treatment: they are considered either fully trapped
(inside the neutrinosphere) or freely streaming (outside the
neutrinosphere), the neutrinosphere being defined by a trapping
density parameter. The latter parameter has been adjusted such that
both codes produce compatible results for all observables shown in
this paper except the neutrino luminosity leading to a value of
$1 \times 10^{12}$~g~cm$^{-3}$. Due to the obvious limitations of the
leakage scheme in \textsc{ACCEPT}, the neutrino luminosity cannot be
well reproduced within this scheme. This does, however, not affect the
results of Secs.~\ref{ss:mass_model} and \ref{s:relevant_nuclei}. We
have checked that the limitation of the neutrino source terms to
charged-current reactions in \textsc{ACCEPT} is irrelevant for the
results presented in those sections, too. Finally, in all the
simulation presented here, initial models (progenitor star) come from
publicly available data computed by Woosley \textit{et
  al.}~\cite{woosley-02}\footnote{\url{https://2sn.org/stellarevolution/}}.

\subsection{Equations of state}\label{ss:eos}

During the different stages of the core collapse
evolution wide domains of density ($10^{-12} \lesssim n_B \lesssim 1$
fm$^{-3}$), temperature ($ 0.1 \lesssim T \lesssim 50$ MeV) and charge
fraction ($ 0.01 \lesssim Y_e \lesssim 0.6$) are explored. Matter
consists of baryons, leptons (electrons, positrons, neutrinos and
antineutrinos) and photons, and it has a homogeneous/inhomogeneous
structure at supra-/sub-saturation densities. Leptons and photons
interact weakly and are usually treated as ideal Fermi and,
respectively, Bose gases. Composition and thermodynamics of baryonic
matter is still under study, because of the
uncertainties related to the effective interactions and difficulties
in the modelling.

The so far most intensively used EoS models
\cite{Lattimer_NPA_1991,Shen_NPA_1998} employ the so-called Single
Nucleus Approximation (SNA). It describes baryonic matter at
sub-saturation densities as a mixture of a uniform distribution of
self-interacting nucleons, a free gas of $\alpha$-particles and a
unique cluster of nucleons, all of which in thermal and chemical
equilibrium with respect to the strong interaction. Interactions
between unbound nucleons and nuclear clusters are included via the
classical excluded volume approximation and in-medium modifications of
cluster surface energy. The shortcoming of this approach becomes
obvious at high temperature, where the macroscopic thermal equilibrium
state corresponds to a collection of distinct microscopic states. The
SNA is known to have only a negligible impact on thermodynamic
quantities \cite{Lattimer_NPA_1985}, but it could affect the weak
interaction rates, highly sensitive to structure effects and thus
finally impact the astrophysical evolution~\cite{Hempel_ApJ_2012}.

A more sophisticated approach consists in employing an extended
Nuclear Statistical Equilibrium (NSE) model which accounts for the
entire nuclear distribution. Several NSE models and resulting EoS have
been proposed in the last decade, see e.g.~\cite{Hempel_NPA_2010,
  Hempel_ApJ_2012, Gulminelli_PRC_2015, Raduta_2019, TN_2017,
  FYSS_2017,Typel_2018}. For the interaction between nucleons,
different relativistic mean field models, Skyrme effective
interactions or variational approaches have been employed that span
significant ranges of nuclear matter parameters in both isovector and
isoscalar channels, accounting thus well for present day uncertainties
in the nuclear matter EoS. Here we are mainly interested in the
infall phase, i.e. matter at sub-saturation densities and with
moderate isospin asymmetries. Consequences of the above uncertainties
in the nuclear interactions on stellar matter are small in this
regime, since a significant amount of matter is bound in clusters and,
by construction, all effective interactions offer a fair description
of ground state nuclei. More important differences among the
different NSE models arise from the modelling of nuclear clusters via
the treatment of i) (thermally) excited states, ii) maximum allowed
isospin asymmetry, iii) nuclear level density and iv) nuclear binding
energies away from the valley of stability, where no experimental data
exist. It has been shown that the latter point strongly influences
the nuclear abundances under the thermodynamic conditions during
collapse~\cite{Raduta2016,Furusawa2017}. Additional differences exist
in the identification of the bound and unbound part of nuclear
clusters to define abundances, but this point did not sizably modify
the EoS \cite{Raduta_2019}. For further details, the reader is
referred to \cite{Hempel_NPA_2010,Raduta_2019}.

For this work, as a fiducial case we will consider the NSE model
of~\cite{Hempel_NPA_2010} with DD2 \cite{DD2} relativistic mean field
effective interaction, see \textcite{Hempel_ApJ_2012}. This model
takes into account the ensemble of nuclei whose masses have been
calculated within the Finite Range Droplet Model in \citet{FRDM}. To
get an idea of EoS effects, we will consider a second model, the NSE
EoS by~\cite{Gulminelli_PRC_2015}. It employs the SLy4 \cite{SLy4}
Skyrme effective interaction. Nuclear clusters have $2 < A <300$ and,
in principle, any value of the isospin asymmetry. Their binding
energies are complementarily given by experimental data
\cite{AME2012a, AME2012b}, the predictions of the 10-parameter model
by Duflo and Zuker \cite{DZ10} and a liquid drop model
parameterization \cite{Danielewicz_NPA_2009}, with parameters
harmonized with SLy4.

Finally, in order to assess the effect of shell closures far from
stability, we will compare different models for the nuclear masses,
using the perturbative method described in \textcite{grams2018} build
upon on the EoS by \textcite{Lattimer_NPA_1991}. Specifically, the
already mentioned phenomenological Duflo and Zuker mass
model\cite{DZ10} (DZ10), which imposes the same magic numbers all over
the nuclear chart, will be compared with the Brussels-Montreal
microscopic mass model\footnote{The mass table for this model is
  available on the BRUSLIB database at
  http://www.astro.ulb.ac.be/bruslib/, see also Ref.~\cite{xu2013}.}
HFB-24 \cite{gcp2013}. The latter, based on the self-consistent
Hartree-Fock-Bogoliubov method, uses a 16-parameter generalized Skyrme
effective nucleon-nucleon interaction with a realistic contact pairing
force, and predicts a considerable quenching of the $N=50$ shell gap
far from stability \cite{gcp2013}. Both DZ10 and HFB-24 provide an
excellent reproduction of measured masses, with a root-mean square
deviation of about $0.5$~MeV with respect to the 2012 Atomic Mass
Evaluation \cite{AME2012b}. We will consider, too, as an
extreme case, masses as described by the Compressible Liquid Drop
Model (CLDM) of the Lattimer-Swesty EoS \cite{Lattimer_NPA_1991},
which completely neglects shell effects.

\subsection{Electron capture rates}\label{ss:ecap_models}

The rate of a generic weak interaction reaction-- electron and
positron capture and $\beta$-decays, depends --apart from physical
constants-- on the nuclear transition strength and a phase space
factor. At finite temperature weak reactions involve several states
in the parent and daughter nuclei, such that nuclear structure effects
enter the transition strength via both, nuclear energy levels and
transition matrix elements. Different reaction channels can contribute
to the latter. In most cases, it is dominated by Fermi (vector) and
Gamow-Teller (axial) contributions. The phase space factor depends on
the electron capture reaction kinematics and takes the form of
integrals over the momenta of incoming and outgoing particles
\cite{FFN_1980}. As such, it shows strong dependence on thermodynamic
conditions.

The first systematic calculation
under stellar conditions is due to Fuller, Fowler and Newman
\cite{FFN_1980,FFN_1982a,FFN_1982b,FFN_1985} who have also made
available \cite{FFN_1982b} the first weak interaction rate table for
nuclei with masses between 21 and 60 on a wide temperature
$T$-electron density $n_e$ grid with $10^7 \leq T \leq 10^{11}$ K,
and $10 \leq n_e \leq 10^{11}$ g/cm$^3$.

The fact that astrophysical simulations require high accuracy weak
interaction rates motivated further extensive microscopic calculations
optimized on experimental data considering in general the same
$T$-$n_e$ grid as \textcite{FFN_1982b}. Different techniques have been
employed, ranging from large scale shell model (LSSM) calculations
\cite{Oda1994, LMP_NPA_2000, LMP_ADNDT_2001} to random-phase
approximation \cite{Paar_PRC_2009, Niu_PRC_2011}, and quasiparticle
random-phase approximation (QRPA) \cite{Nabi_1999, Nabi_2004}.  By
accounting for all possible correlations among valence nucleons in a
major shell, LSSM calculations offer the most accurate microscopic
description available to date, as testified by its ability to
reproduce the measured GT distributions, lifetimes and low energy
spectroscopy \cite{Caurier_1999}. They exist for $sd-$
($17 \leq A \leq 39$) \cite{Oda1994} and $fp-$shell nuclei
($45 \leq A \leq 65$) \cite{LMP_NPA_2000, LMP_ADNDT_2001}. The mass
domain $65 \leq A \leq 80$ is covered by the table of
Ref. \cite{Pruet2003}, which employs an empirical approach. Finally
weak interaction rate tables for $sd$- , $fp$- and $fpg$-shell nuclei
with $18 \leq A \leq 100$ are given in \cite{Nabi_1999,Nabi_2004},
which employ QRPA.  \ar{QRPA has been also recently employed in \cite{Titus2019} for
calculating EC rates of neutron rich nuclei with $26 \leq Z \leq 41$ and
$75 \leq A \leq 93$, which correspond to the high sensitivity region of
Ref.\cite{Titus2017}. Hybrid models which use shell-model
  Monte-Carlo (SMMC) \cite{Dean_PRC_1998} or Fermi-Dirac
  parametrizations \cite{Juodagalvis_NPA_2010} to determine the
  population of excited states and RPA techniques for weak interaction
  rates have also been proposed and exploited to extend the existing
  data to heavier and more neutron rich nuclei. In this way, the work
  by \textcite{Juodagalvis_NPA_2010} contains information for nuclei
  with $66 \leq A \leq 120$ (250 nuclides) and with
  $28 \leq Z \leq 70$ and $40 \leq N \leq 160$ (2200 nuclides).  In
  particular, cross shell correlations which are important to overcome
  the $N$=40, 50 and 82 shell gaps are accounted for, in agreement
  with the results of finite temperature SMMC. Moreover, based on the
  observation that the electron Fermi energy grows faster with core
  density than the nuclear $Q$-value, \textcite{Juodagalvis_NPA_2010}
  define a strategy to describe electron captures by a hierarchy of
  nuclear models. They provide NSE-averaged EC rates along two
  collapse trajectories. These rates are not appropriate for other
  studies where the nuclear distribution is calculated consistenly
  from the employed EoS. }

It is easy to see that the available databases cover a
finite mass domain and an isospin asymmetry range close to the valley
of stability. Although strong structure effects translate into EC
rates that, for low temperatures and electron density, can vary by
more than one order of magnitude between neighboring nuclei in the
isotopic chart, the need of estimates for other nuclei, including the
neutron-rich ones copiously populated during the late collapse stages,
and/or thermodynamic conditions out of the grids lead to the use of
extrapolations and approximations within simulations.

\begin{figure}
\tikzsetnextfilename{fig1}
\input{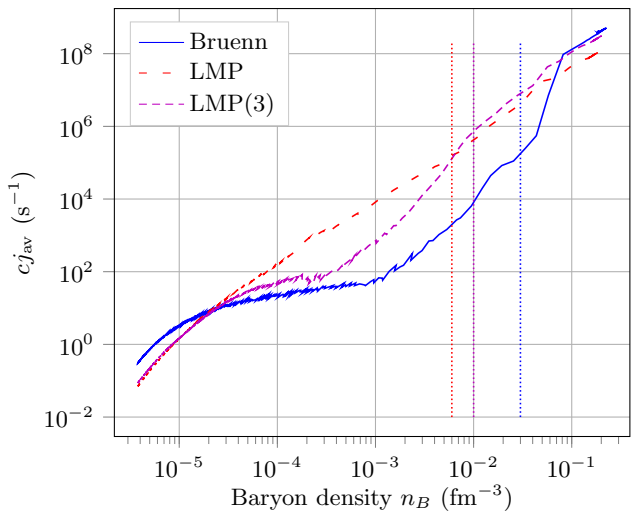}
\caption{(color online) EC rate evolution (labelled by baryon
  density) in the central grid cell during infall using different
  prescriptions for the EC rate on nuclei, see text for details. The
  vertical dashed lines show the density above which
  $\beta$-equilibrium sets in.
  \label{fig:ratesnb}}
\end{figure}

The first parameterization, proposed by~\textcite{Bruenn_1985}, relies on
the independent particle model and estimates the GT matrix element by
the number of
protons in the $\pi 1f_{7/2}$ shell and the number of neutron
holes in the $\nu 1f5/2$ one. The reaction $Q$-value entering the phase
space factor is approximated by the difference between proton and
neutron chemical potential.
This prescription results in the total suppression of EC on
both light and/or neutron-rich nuclei, which is certainly
unrealistic.

\ar{ An improved parameterization, which is presently the most
  extensively used in CCSN simulations, was proposed in
  \textcite{Langanke_PRL_2003}.  It is based on results of SMMC
  calculations at finite temperature in the full $pf-sdg$ shell with
  residual pairing plus quadrupole interactions and RPA calculations
  of EC for nuclei with $65 \leq A \leq 112$.  At variance with the
  independent particle model, all nuclei studied in
  \textcite{Langanke_PRL_2003} manifest holes in the $pf$ shell and,
  for $Z>30$, non-vanishing proton occupation numbers in the $sdg$
  orbitals. This means that GT transitions are unblocked and EC rates
  take place.  This new parametretization reads:}
\begin{eqnarray}
j^{EC}=\frac{\ln 2 \cdot {\mathcal B}}{K} \left(\frac{T}{m_e c^2} \right)^5
\left[ F_4(\eta)-2 \chi F_3(\eta)+\chi^2 F_2(\eta)
\right]~.
\label{eq:lEC}
\end{eqnarray}
In this expression, $\chi=(Q-\Delta E)/T$, $\eta=\chi+\mu_e/T$, where
$Q$ denotes the EC reaction heat, $Q=M(A,Z) - M(A,Z-1)$, with $M(A,Z)$
the nuclear mass, and $\Delta E=E_f-E_i$. $m_e$ and $\mu_e$ stand for
electron rest mass and chemical potential, respectively.  $F_i(\eta)$
denotes the relativistic Fermi integral,
$F_i(\eta)=\int_0^{\infty} dx x^k/(1+\exp(x-\eta))$. ${\mathcal B}$
represents an average value for the nuclear matrix element.  The
constant values proposed in Ref.~\cite{Langanke_PRL_2003},
${\mathcal B}=4.6$ and $\Delta E=2.5$ MeV, are obtained from a fit of
\ar{SMMC+RPA calculations for nuclei with $65 \leq A \leq 112$}, and
shown to correctly reproduce their gross features for the
thermodynamic conditions explored by the central element before bounce
\cite{Langanke_PRL_2003}.  \ar{For lower values of temperature and
  electron density, Eq.~(\ref{eq:lEC}) may nevertheless lead to both
  underestimations and overestimations of microscopic calculations, as
  showed in Refs. \cite{Langanke_PRL_2003,Juodagalvis_JPG_2007}.}
Note that the parameter $\Delta E$ accounts for possible transitions
from and to excited states in the parent/daughter nucleus, so that it
is not necessary to introduce an effective $Q$-value such as in
\textcite{Sullivan16}.

The above parameterization, Eq.~(\ref{eq:lEC}), being integrated over
neutrino energies, can actually not directly be implemented into our
multigroup neutrino treatments. Instead we use the following
expression for the neutrino creation rate ($\Theta$ is the usual
Heaviside step function and $f_{e^-}$ the electron distribution function):
\begin{align*}
j(\epsilon)&=\Theta(\epsilon-\chi T)\frac{\ln 2 \cdot {\mathcal B}}{K} \left(\frac{1}{m_e c^2} \right)^5 \frac{(hc)^3}{4\pi c} \\ &\qquad \times   (\epsilon-\chi T)^2  f_{e^-}(\epsilon-\chi T) \numberthis{}
~.
\label{eq:lEC_fmt}
\end{align*}
which once integrated over neutrino energy $\epsilon$ assuming a
vanishing neutrino distribution function exactly reproduces
Eq.~(\ref{eq:lEC}). Throughout this work this parameterization will be
denoted LMP(0).
\begin{figure}
\tikzsetnextfilename{fig2}
\input{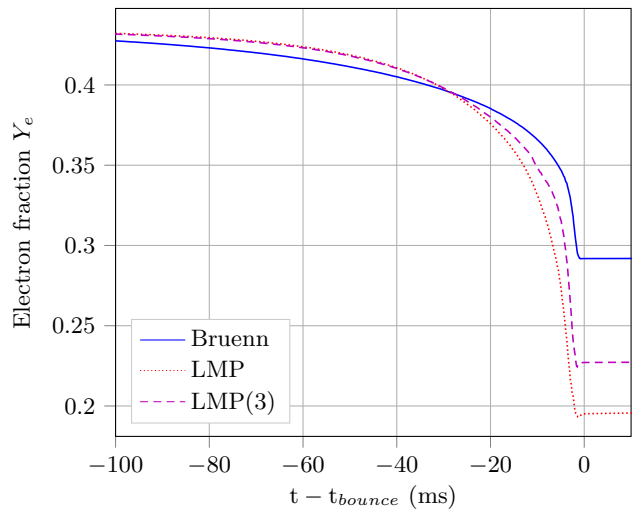}
\caption{(color online) Time evolution of the electron fraction $Y_e(r=0)$
  for the central element of the numerical grid for all three EC
  models during the late pre-bounce phase.
  \label{fig:ye_c_t}
}
\end{figure}

Eq.~(\ref{eq:lEC}) has been recently generalized by \textcite{Raduta2017},
allowing for a temperature, electron density, and isospin
$I=(N-Z)/A$-dependence as well as for odd-even effects in $\Delta E$.
The temperature dependence contains two competing effects: the
increasing number of excited states in the daughter nucleus with
increasing temperature, leading to a larger $\Delta E$ and the
decreasing electron chemical potential, $\mu_e$, for fixed electron
density, which shows the opposite trend. With increasing $n_e$, more
excited states and higher energies can be populated and thus
$\Delta E$ increases, too. Finally the isospin dependence and
odd-even effects of $\Delta E$ are introduced to account for nuclear
structure effects in the centroid of the GT resonance, as computed
within LSSM for $pf$-shell nuclei \cite{LMP_NPA_2000}.

As shown in Ref. \cite{Raduta2017} for a wide range of thermodynamic
conditions typical to late stage evolution of core-collapse, the most
important improvement of Eq.~(\ref{eq:lEC}) arises from isospin and
odd-even effects, i.e. nuclear structure properties. These features
are particularly useful in accounting for the large dispersion of
$j^{EC}$ in a given $Q$-value bin -- as shown by the data in
\cite{Oda1994, LMP_ADNDT_2001, Pruet2003, Nabi_1999, Nabi_2004}--, and
for the extrapolation to large negative $Q$, typical to intermediate
mass neutron-rich nuclei \cite{Raduta2017}. In relation with the
latter effect we point out that, for $T \gtrsim 1$ MeV and the highest
$n_e$-values considered in the grid, the improvements discussed in
\textcite{Raduta2017} lead to EC rates lower by two orders of
magnitude or more than those produced assuming
Eq.~(\ref{eq:lEC}). Given the overall neutron-enrichment of stellar
matter before bounce, reduction of two orders of magnitude of the
individual rates on neutron-rich nuclei entails a reduction of up to
one order of magnitude for the average EC rate summed over the
complete nuclear distribution~\cite{Raduta2017}. Further
developments, which are out of the scope of present work, should also
account for isospin effects on the GT strength \cite{LMP_NPA_2000} and
temperature-dependent Pauli blocking.

Out of the different improved versions of Eq.~(\ref{eq:lEC}) proposed in
\cite{Raduta2017} we shall here consider only one, model 3, which best
reproduces microscopic data.
To be more precise, instead of employing a global value $\Delta E$, for each
grid point of $n_e(i), T(j)$ in the microscopic calculations we write
\begin{equation}
\Delta E^{(AB)} (n_e(i), T(j)) = b_{i,j}^{(AB)} I + c_{i,j}^{(AB)}~,
\label{eq:lmp3}
\end{equation}
and intermediate values of $n_e,T$ are obtained by linear
interpolation. The coefficients $b,c$ have been determined in
\textcite{Raduta2017} by a least square fit to LSSM
calculations. Eq.~(\ref{eq:lmp3}) assumes a linear isospin dependence
and odd even effects are included by employing different coefficients
$\{(AB)\} = \{(OO),(OE),(EE)\}$ for odd-odd, odd-even and even-even
nuclei, respectively. For details and in particular values of the
coefficients, see the appendix of \textcite{Raduta2017}. Hereafter
this parameterization, implemented in its energy dependent form
Eq.~(\ref{eq:lEC_fmt}), will be referred to as LMP(3).

During the advanced stage of the collapse, before $\beta$-equilibrium
is reached at the center of the star, $T$ and $n_e$ exceed the values
covered by the weak rate tables and thus $\Delta E$ cannot be fitted
in that region. We have used two different ways to extrapolate
$\Delta E$ in this case: (i) first order extrapolation or (ii)
constant value fixed to the last available $T, n_e$ grid point. The
predicted EC rate (and subsequently the evolution of $Y_e$) differ by less
than 3.5~\% between these two procedures, indicating
that EC rates under those extreme conditions are of little relevance
and that the extrapolation procedure for $\Delta E$ is not of great influence.

\ar{
  A word of caution has to be added.
  Being based on fits of $pf$-shell nuclei, the approximation proposed in
  Ref. \cite{Raduta2017} might not be appropriate for EC rates of heavy nuclei,
  which, due to cross-shell correlations, manifest suppression of Pauli
  blocking effects \cite{Zhi_NPA_2011}.
  This unblocking of the GT transition was predicted by theoretical models
  for nuclei with $Z<40$ and $N>40$, and confirmed by experiments.
}
\section{Evolution of the collapse}\label{s:results}

\begin{figure}[h]
\tikzsetnextfilename{fig3}
\input{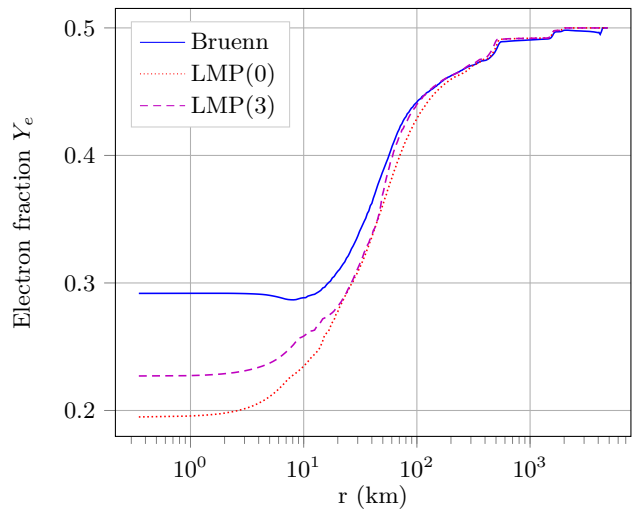}
\caption{(color online) Radial profiles of the electron fraction at bounce
  employing the three different EC rate prescriptions.
  \label{fig:yebounce}
}
\end{figure}

All simulations start from an unstable stellar model as mentioned in
Sec.~\ref{ss:coconut}. Except for Sec.~\ref{sss:progenitors}, where we
study the influence of the progenitor model on the results, a
15~$M_\odot$ progenitor from \textcite{woosley-02} will be used, the
\texttt{s15} model, see Table~\ref{tab:progenitors}. We then follow the
collapse of the iron core, with establishment of $\beta$-equilibrium
at the center and finally stop the simulation a few milliseconds after
bounce. Simulations in Sections \ref{ss:ec_rates} to \ref{sss:eos}
have been performed with the \textsc{CoCoNuT} code, whereas those in
Sections \ref{ss:mass_model} and \ref{s:relevant_nuclei} with the
ACCEPT code.

\subsection{Influence of electron capture rates}
\label{ss:ec_rates}
Our fiducial simulation starts from a $15\ M_\odot$ progenitor
(zero-age main-sequence) labelled \texttt{s15} in the catalog by
Woosley \textit{et
  al.}~\cite{woosley-02}\footnote{\url{https://2sn.org/stellarevolution/}}. As
a first test we plot in Fig.~\ref{fig:ratesnb} averaged EC rates in
the central cell of our numerical grid, as functions of the baryon
density in that cell during infall, for all three approaches detailed
in Sec.~\ref{ss:ecap_models}: the original one by
Bruenn~\cite{Bruenn_1985}, LMP(0)~\cite{Langanke_PRL_2003} and
LMP(3)~\cite{Raduta2017}. To average the rates, we assume a
Fermi-Dirac equilibrium distribution for neutrinos,
\begin{eqnarray}
j_{\mathrm{av}} = \int_0^\infty j(\epsilon) \epsilon^3 f_\nu^{(eq)}(\epsilon) d\epsilon ~.
\label{eq:ec_av}
\end{eqnarray}

The density where $\beta$-equilibrium is achieved with the different
EC prescriptions is shown by dashed lines. From this figure, it is
obvious that up to a baryon density of
$n_B \simeq 2 \times 10^{-5} \textrm{ fm}^{-3}$, EC rates given by
Bruenn's model are higher than the LMP ones. The behavior gets
inverted at densities between $2 \times 10^{-5} \textrm{ fm}^{-3}$ and
$7\times 10^{-4} \textrm{ fm}^{-3}$ because, within this density
interval, many neutron-rich nuclei are populated with vanishing EC
rates in the simplified Bruenn's approach. This density region is
located close to the onset of $\beta$-equilibrium with the highest EC
rates, such that it is to be expected that the difference in EC rates
between the three models is relevant for the evolution. The
importance of EC on neutron-rich nuclei has already been noted in
Ref.~\cite{Hix2003}, comparing simulations employing LMP(0) rates and
Bruenn's rates.

\begin{figure*}
\tikzsetnextfilename{fig4}
\input{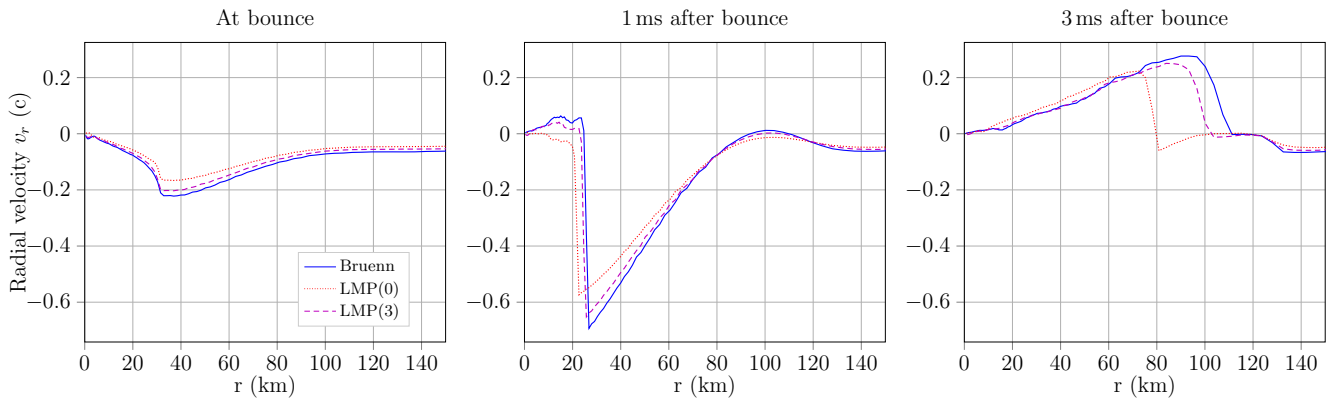}
\caption{(color online) Comparison of shock propagation at different
  instants during the early post-bounce phase with the three different
  EC rate prescriptions.
  \label{fig:v1ebounce}
}
\end{figure*}
The electron fraction $Y_e$ is directly linked to the EC rates. Its
time evolution at the center of the star is plotted in
Fig.~\ref{fig:ye_c_t}. During infall, the model with Bruenn's rates
shows a different behavior with respect to the two LMP models. This
can be understood as follows: during most of the collapse, the central
baryon density is lower than $2\times 10^{-5} \textrm{ fm}^{-3}$ and
EC rates by Bruenn are higher than the other ones (see
Fig.~\ref{fig:ratesnb}), which implies a lower $Y_e$ for the model
using Bruenn's EC rates before $t\simeq 210 \textrm{ ms}$. As shown in
Fig.~\ref{fig:ratesnb}, for higher densities until the onset of
$\beta$-equilibrium, the situation is inverted and EC rates by LMP(0)
and LMP(3) models are higher, leading to a stronger decrease of $Y_e$
after $t\simeq 210 \textrm{ ms}$. This behavior shows again the
importance of EC rates in the density range
$2\times 10^{-5} \textrm{ fm}^{-3} \leq n_B \leq 7 \times 10^{-4}
\textrm{ fm}^{-3}$
for the evolution of the collapse, where EC on neutron-rich nuclei
occurs. At bounce differences of about (30\%) in $Y_e(r=0)$ are
observed, which can be explained by the fact that EC in this density
region occurs predominantly on nuclei for which no microscopic
calculations exist~\cite{Sullivan16}. Although qualitatively the
behavior is very similar between the two LMP models, the decrease in
$Y_e$ is less pronounced employing model LMP(3).

The study of Sullivan et al~\cite{Sullivan16}, where -- in order to
account for uncertainties in EC rates -- these are scaled by a
constant factor of 10, indicates differences of up to $\pm 30\%$ in
the central $Y_e$ at bounce. The nuclear physics considerations
(mainly isospin dependence and odd-even effects, see
Section~\ref{ss:ecap_models}) entering the improved model LMP(3)
considerably reduce the EC rates in the late stages of collapse and
thus clearly point to a higher $Y_e$ at bounce than for LMP(0).

In Figure~\ref{fig:yebounce} we display the radial $Y_e$ profiles at
bounce for the three EC rate models. These results indicate
that differences appear mostly in the central region of the collapsing
star, i.e. for $r \lesssim 50 \textrm{ km}$ at bounce, where
neutron-rich nuclei are most abundant. As the three EC rate models
presented in this paper mostly differ on these exotic nuclei, the
overall behavior displayed in Fig.~\ref{fig:yebounce} is
understandable.

Going further, we now look at the influence of our EC rate models on
the collapse and bounce dynamics. Since the collapse can be seen
essentially as a free fall, it is clear that the collapse time shows
only little difference between the models: it is about 4\% larger with the
two LMP models than with Bruenn's rates. This small difference can be
understood from the fact that in the early collapse phase, electron
degeneracy pressure is dominant. The lower electron fraction for the
model with Bruenn's rates (see Fig.~\ref{fig:ye_c_t}) in this phase thus
explains the accelerated collapse.

Radial velocity profiles at bounce, and at two instants after bounce
($1 \textrm{ ms}$ and $3 \textrm{ ms}$) are shown in
Fig.~\ref{fig:v1ebounce}. Although the radius at which the shock is
formed is the same for all three EC rate models (left panel), the
situation is different 3~ms after bounce (right panel), where the
shock is seen to have reached the largest distance from the center for
the Bruenn case, and closest to the center with the LMP(0) rates. This
can be interpreted in terms of the inner core mass, i.e. the mass of
the matter inside the shock formation radius: the higher the mass of
this inner core, the larger the kinetic energy given to the
shock. Additionally, if the inner core mass is larger, the iron layers
that must be crossed by the shock are thinner, thus making the shock
lose less energy. Values of the inner core mass at bounce are computed
to be $0.31\, M_\odot$, $0.4\, M_\odot$ and $0.45\, M_\odot$, for models
with EC rates from LMP(0), LMP(3) and Bruenn, respectively, confirming
the above reasoning. Please note that, if we had shown the shock
position as a function of the enclosed mass and not as a function of the
radius (left panel of Fig.~\ref{fig:v1ebounce}), the difference in the
inner core mass at bounce would have induced visible differences, see
e.g. Fig.~7 of \textcite{Sullivan16}.

The ordering of the inner core mass can in turn be
understood as a consequence of the electron fraction evolution
discussed above: The mass of the inner core at bounce is roughly
proportional to $\langle Y_{L^{(e)}}^2\rangle$, the mean fraction of
trapped leptons squared~\cite{BurrowsLattimerYahil} which is fixed and
given by $Y_e$ at the moment when neutrino trapping sets in.

\begin{figure}
\tikzsetnextfilename{fig5}
\input{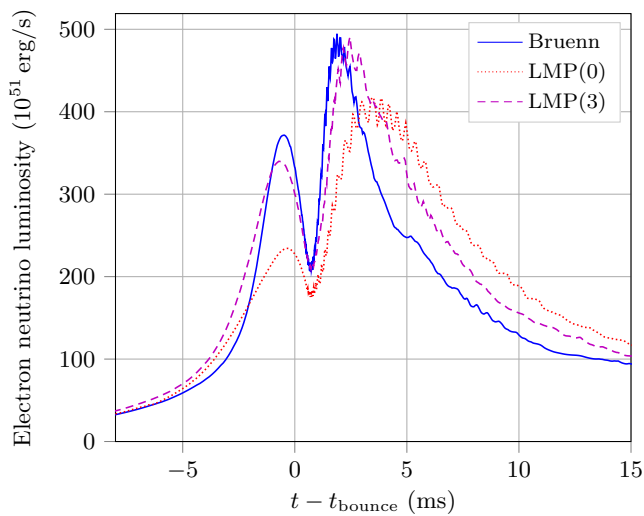}
\caption{(color online) Time evolution of electron neutrino luminosity
  around bounce, for the three different EC rate prescriptions.
  \label{fig:luminosity}
}
\end{figure}

\begin{figure*}
\tikzsetnextfilename{fig6}
\input{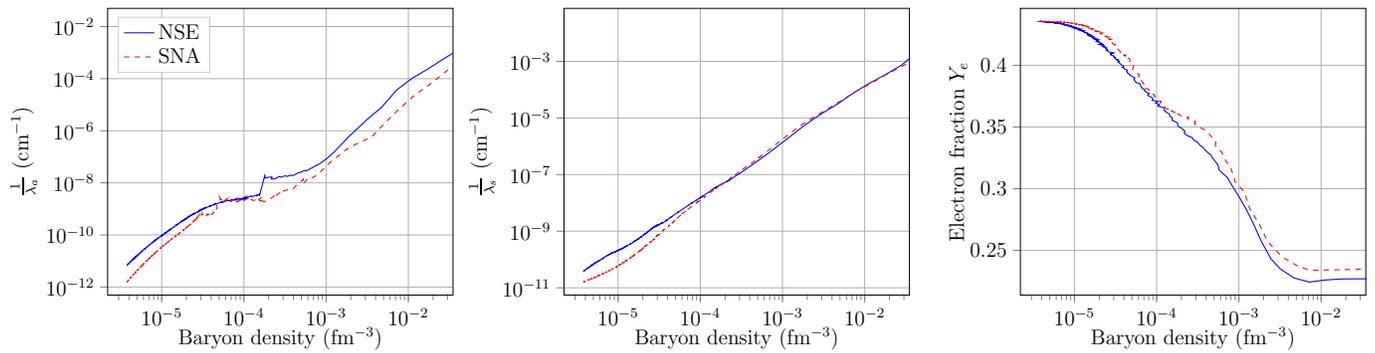}
\caption{(color online) Neutrino inverse mean
  free path as functions of baryon density in the central numerical
  cell with reactions computed either on a single mean nucleus (SNA) or
  on a statistical ensemble of nuclei (NSE): contribution of
  electron capture effects only (left panel) and scattering effects only
  (central panel). Electron fraction time evolution for both models (right panel).
  \label{fig:sna_nse}
}
\end{figure*}

Finally, in Fig.~\ref{fig:luminosity} are drawn, for each EC rate
model, the corresponding electron neutrino luminosity as functions of
time. Qualitatively, the behavior is similar: for all curves one notes
first an increase before bounce corresponding to the increase of
electron captures. About $1 \textrm{ ms}$ before bounce, the density
increases so that neutrinos are trapped and luminosity starts to
decrease. A few milliseconds after bounce a second peak appears, due
to the shock reaching the neutrinosphere.

Quantitatively, the first peak is strongest for Bruenn's rates,
reflecting the higher EC in the early collapse phase for that model.
Following Ref.~\cite{Sullivan16}, the ordering of the second peak,
highest and fastest for Bruenn, followed by LMP(3) and LMP(0), can be
interpreted as consequence of the shock propagation velocity. If the
shock propagates faster, it simply reaches the neutrinosphere
earlier. The velocity profiles given in Fig.~\ref{fig:v1ebounce}
confirm this interpretation. The total energy taken away by electron
neutrinos, up to $15 \textrm{ ms}$ after bounce can be computed and is
roughly independent of the EC rate model: $4.58\times10^{51}$~ergs
with Bruenn's model, $4.90\times 10^{51}$~ergs with LMP(0) and
$4.99\times 10^{51}$~ergs with LMP(3).

\subsection{Influence of other parameters}

\subsubsection{SNA vs NSE}
We study here the influence of other parameters on the infall
evolution, starting with a comparison between NSE or SNA approaches
within the EoS (see Sec.~\ref{ss:eos} for details) when computing
neutrino reactions. To that end,
we simulate infall using EC rates
computed with the LMP(3) model, the \texttt{s15} progenitor model as well as
the HS(DD2) EoS. SNA calculations thereby extract the average nucleus
from the entire available NSE distribution. Thermodynamic quantities
are thus unchanged between SNA and NSE and we can more easily isolate
neutrino reaction effects. \ar{In the SNA case, the Q-value needed for both LMP models is computed using the nucleus obtained by rounding off the avagerage $(A,Z)$ to the closest integers.}

\ar{This approach differs from previous studies comparing SNA and NSE
  presciptions. Studies focussing on thermodynamic quantities of
  course recalculate the full EoS in SNA or NSE approach, respectively,
  but employ in general simplified neutrino treatments, see
  e.g. \textcite{Hempel_ApJ_2012}, who calculate the NSE weak
  reactions extracting two average nuclei from the full distribution
  with Bruenn rates. They confirm that considering the full nuclear
  distribution for thermodynamic quantities has only a minor
  impact~\cite{Burrows_84, Hempel_ApJ_2012} on core collapse. On the
  contrary, the importance of taking into account the full nuclear
  distribution for calculating weak rates is well known, see
  e.g. \cite{Hix2003}. Our approach allows to properly investigate the
  issue since we consistently compute weak rates from the nuclear
  distribution of the underlying EoS.}

The left panel of Fig.~\ref{fig:sna_nse} shows a comparison of
averaged neutrino inverse mean free paths from EC processes in the
central cell,
\begin{eqnarray}
  \frac{1}{\lambda_a} = \int_0^\infty \kappa_a^*(\epsilon) \epsilon^3 f_\nu^{(eq)}(\epsilon) d\epsilon \left/ \int_0^\infty \epsilon^3 f_\nu^{(eq)}(\epsilon) d\epsilon \right.
\label{eq:mfp_av}
\end{eqnarray}
as function of baryon density. $\kappa_a^*$ denotes here the
absorption opacity corrected for stimulated absorption given by
$j(\epsilon)/f_\nu^{(eq)}$~\cite{Bruenn_1985}. This Fermi-Dirac
weighted average yields the correct mean free path for gray energy
transport in an optically thin medium. As expected, most important
differences appear in the density ranges above $ n_B \gtrsim 10^{-4}
\mathrm{fm}^{-3}$ where the nuclear distribution is large and
potentially dominated by more than one peak.

The middle panel of Fig.~\ref{fig:sna_nse} shows a comparison of NSE
and SNA for the averaged inverse neutrino mean free path obtained from
scattering off nuclei. We assume isoenergetic scattering and include
corrections due to ion correlations and electron screening, see
\textcite{Horowitz_1997} and \textcite{Bruenn_1997} for detailed
expressions. The right panel shows the time evolution (labelled by
baryon number density) of $Y_e$ in the central cell for both
cases.

It is obvious that, although differences between SNA and NSE occur
mainly in the region where the nuclear distribution is large, the
overall effect is much smaller than using different prescriptions for
EC rates. We should stress, however, that Bruenn's rates as well as
both parameterizations LMP(0) and LMP(3) average over nuclear
structure effects and the difference between NSE and SNA might become
more important when employing microscopic rates all over the nuclear
chart, which are presently not available.

\subsubsection{EoS dependence}
\label{sss:eos}
The EoS model can have some influence on the electron fraction,
too. To check this, we compare simulations obtained with both
extended NSE EoS models described in Sec.~\ref{ss:eos}, HS(DD2) from
\textcite{Hempel_ApJ_2012} and the SLy4 from \textcite{Raduta_2019}
using model LMP(3) for the EC rate. $Y_e$ at the star's center shows
little differences during the collapse, and almost no difference in
the resulting value at onset of $\beta$-equilibrium and at bounce. For
the evolution of the central density after bounce, only small
differences between both models were noted, too. The only noticeable
discrepancy between the two EoS models could be seen in the
temperature evolution at the center of the star: the EoS by
\textcite{Raduta_2019} always leads to slightly lower values than the
one by \textcite{Hempel_ApJ_2012}, but these differences have little
influence on the overall dynamics during pre-bounce and early
post-bounce.

\ar{The EoS dependence of the CCSN evolution before and after bounce
  has been previously considered in
  Refs. \cite{Sumiyoshi_ApJ_2005,Hempel_ApJ_2012,Fischer_EPJA_2013,Sullivan16,Nagakura2019},
  which employed a relatively wide collection of models in both SNA
  and NSE frameworks.  The unanimous conclusion is that a certain,
  though limited, dependence is observed for practically all
  considered thermodynamic and dynamic quantities in all stages of the
  evolution as well as for the deleptonisation rate, $Y_e$ and
  neutrino signals.  \textcite{Fischer_EPJA_2013} reaches the
  conclusion that $Y_e$ of the protoneutron star and its evolution
  depend on the symmetry energy.  }
\begin{table}[t]
  \squeezetable
  \begin{tabular}{|c|c|c|c|}
    \hline
    Progenitor name & Metallicity & ZAMS mass & Mass at
                                                collapse\\
    \hline
    \texttt{s15} & Solar & $15\, M_\odot$ & $2.1\, M_\odot$ \\
    \texttt{s25} & Solar & $25\, M_\odot$ & $2.9\, M_\odot$ \\
    \texttt{s40} & Solar & $40\, M_\odot$ & $2.6\, M_\odot$ \\
    \texttt{u15} & $10^{-4}\times$Solar & $15\, M_\odot$ & $2.0\, M_\odot$ \\
    \texttt{u25} & $10^{-4}\times$Solar & $25\, M_\odot$ & $2.3\, M_\odot$ \\
    \texttt{u40} & $10^{-4}\times$Solar & $40\, M_\odot$ & $4.6\, M_\odot$ \\
    \hline
  \end{tabular}
  \caption{Progenitor models taken from \textcite{woosley-02}. Mass at
    collapse demotes the mass present on the numerical grid at the
    beginning of simulation.}\label{tab:progenitors}
\end{table}

\subsubsection{Progenitor dependence}
\label{sss:progenitors}
We have also explored the role of the progenitor model in the
determination of electron fraction at bounce by considering six
different models from \textcite{woosley-02}, see
Tab.~\ref{tab:progenitors}. The simulations employ LMP(3) EC rates and
the HS(DD2) EoS. Although the overall collapse time may noticeably
depend on the type of progenitor with differences of up to 25\%, the
electron fraction at bounce varies only from $0.23$ to $0.27$, with
the exception of the model \texttt{u40}, for which it drops down to
$0.2$. This last point can be understood from the large mass present
on the numerical grid at the beginning of the collapse (see
Tab.~\ref{tab:progenitors}). We thus confirm conclusions from
\textcite{Sullivan16}, who showed that the detailed progenitor model
can have less influence on the electron fraction at bounce than the
precise EC rate prescription.


\begin{figure}
    \resizebox{\columnwidth}{!}{\input{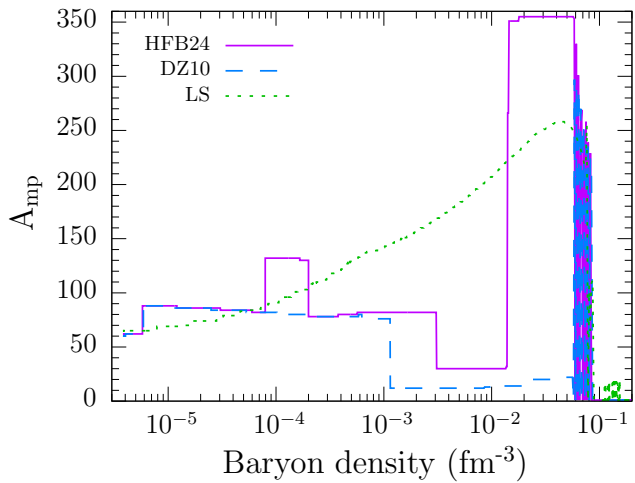}}
\caption{(color online) Baryonic number of the most probable nucleus
  in the central element during infall as function of time (labelled
  by the baryon number density). Three different mass models are
  considered.
  \label{fig:time_A}
}
\end{figure}

\begin{figure*}
   \resizebox{0.3\textwidth}{!}{\input{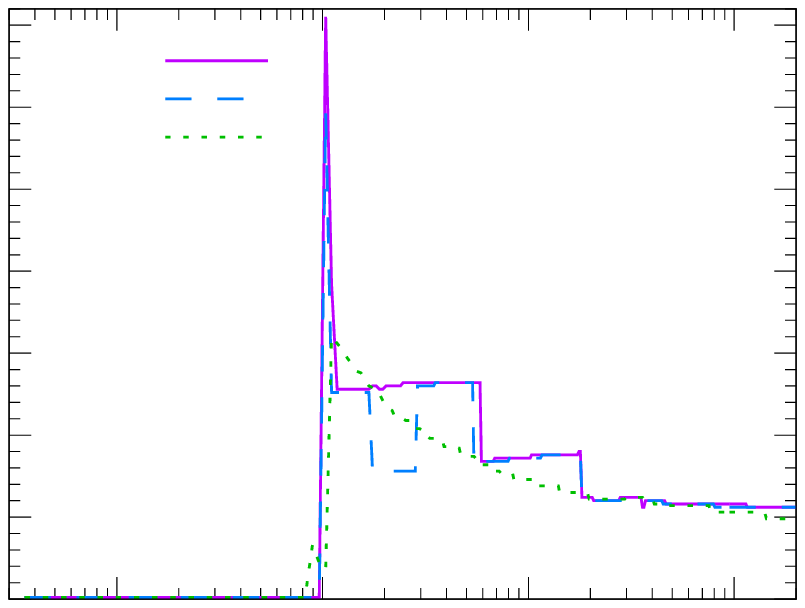}} \hfill  \resizebox{0.3\textwidth}{!}{\input{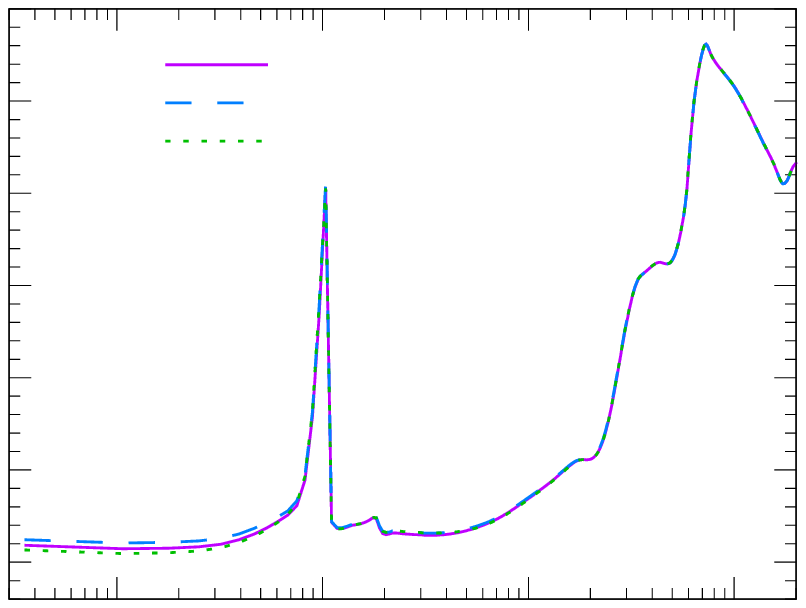}} \hfill \resizebox{0.3\textwidth}{!}{\input{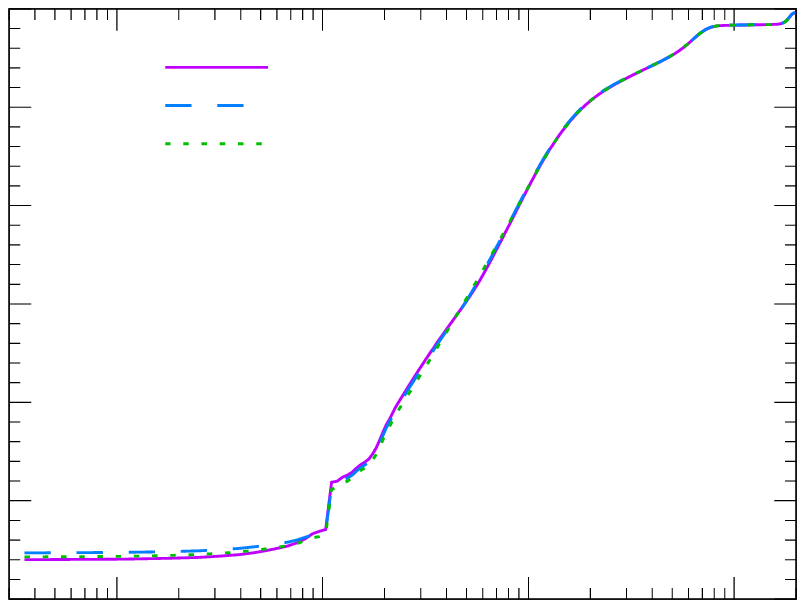}}
\caption{(color online)  Radial profile at the time of bounce of the baryonic number of the most probable
nucleus  (left), the entropy (central) and the electron fraction (right), using three different mass models.
  \label{fig:mass_model}
}
\end{figure*}

\subsubsection{Influence of the nuclear mass model} \label{ss:mass_model}

Finally, we have examined the influence of the nuclear mass model on
the dynamics of the collapse. Previous works \cite{Raduta2016,
  Furusawa2017, grams2018} have shown that the composition of matter
in the thermodynamic conditions of core collapse considerably varies
according to the functional used for extrapolating nuclear masses
beyond known ones. In particular, since the composition is dominated
by magic nuclei \cite{Sullivan16}, it was reported in
Refs.~\cite{Raduta2016,Furusawa2017,grams2018} that a potential
modification of magicity far from stability would strongly affect the
distribution of matter.

For the present study, we have used the perturbative method introduced
in Ref.~\cite{grams2018} to compute the NSE distribution starting from
a given density functional for the EoS. As mentioned in
Sec.~\ref{ss:eos}, the Lattimer-Swesty functional has been applied for
that purpose and simulations have been performed with the
\textsc{ACCEPT} code using LMP(0) parameterization for the EC rates.

In Fig.~\ref{fig:time_A} the evolution of the most probable nucleus as
a function of time in the central element is shown comparing the
predictions from a CLDM prescription (LS~\cite{Lattimer_NPA_1991}),
with that of two microscopic mass models, DZ10~\cite{DZ10} and
HFB-24~\cite{gcp2013}. In very good qualitative agreement with
Refs.~\cite{Raduta2016,grams2018}, we can see that the presence (in
DZ10 and HFB-24) or absence (in LS) of shell effects impacts in a
considerable way the composition of matter. The HFB-24 model, which
predicts a stronger quenching of the shell gaps far from stability
than the DZ10 model, naturally predicts a faster evolution towards
heavier nuclei. This is expected, since a quenching of the shell gap
reduces the waiting-point effect due to magicity, well known in the
framework of r-process calculations \cite{pearson96}.

\begin{figure}
  \includegraphics[width=\columnwidth]{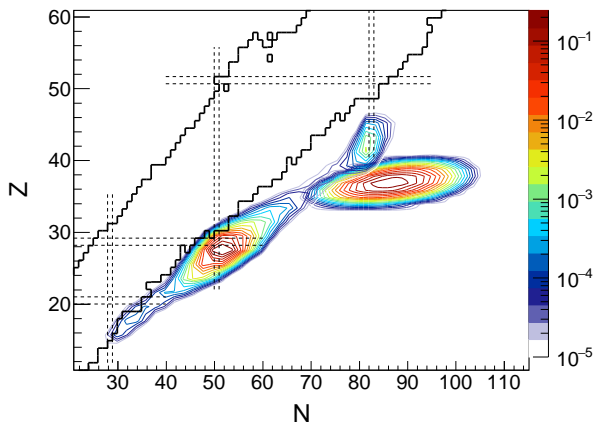}
  \caption{(color online) Distribution of nuclei $(N,Z)$ for a chosen
    thermodynamic condition during the collapse before trapping:
    $T = 1$~MeV; $ \rho_B= 6.02\times 10^{11}$~g~cm$^{-3}$;
    $Y_e = 0.27$. Contour lines correspond to the cluster normalized
    probabilities (red to blue, more to less probable) for the
    original LS model (oval-shaped contour) and for the HFB-24 nuclear
    mass model. See text for details.
  \label{fig:distr-6d11}
}
\end{figure}

However, the differences between the mass models only marginally
affect the global dynamical evolution of the collapse. This is
demonstrated in Fig.\ref{fig:mass_model}, which gives as an example
the radial profile at bounce of different representative
quantities. The behavior of the most probable cluster (left part)
follows the trends already observed in Fig.~\ref{fig:time_A}. Since a
potential magicity quenching does not change the global pattern of
nuclei produced, but only the time at which they appear, it is not
surprising that the profiles at bounce of the different models are
very similar. Less expected is the fact that the electron fraction and
entropy profile (central and right part of Fig.~\ref{fig:mass_model})
of the different models are indistinguishable, meaning that the time
integrated effect of the different compositions is very small. This is true
even for the simplistic liquid drop model (LS) which does not account
for any structure effect, and even if the global distribution of
nuclei is very different between the LS and the other models at all
times (see Fig.~6 of Ref.~\cite{grams2018}). This is clearly
noticeable in Fig.~\ref{fig:distr-6d11} (see also Fig.~6 of
Ref.~\cite{grams2018})\footnote{The nuclear distributions shown in
  Fig.~\ref{fig:distr-6d11} appear quite different from those shown in
  Fig.~6 of Ref.~\cite{grams2018}. This is because in
  Ref.~\cite{grams2018}, the cluster probabilities have been
  calculated on a fixed core-collapse trajectory using Bruenn's rates
  and with a trapping density fixed to $3 \times 10^{11}$~g~cm$^{-3}$,
  while here the calculations have been consistently done in the
  core-collapse simulations using LMP(0) rates and a trapping density
  fixed at $10^{12}$~g~cm$^{-3}$, thus yielding different
  thermodynamic conditions.}, where we plot the distribution of nuclei
for a given thermodynamic condition reached during the collapse in the
center of the star ($T = 1$~MeV; $ \rho_B= 6.02 10^{11}$~g~cm$^{-3}$;
$Y_e = 0.27$) for the LS model (oval-shaped contours) and the HFB-24
mass model (bimodal contours). While for the LS model the most
probable nucleus is located around $N \approx 86$ and $Z \approx 37$,
for the HFB-24 mass model, the most probable nucleus is still located
around the magic number $N=50$ and $Z=28$, and the probabilities show
a bimodal distribution with a second peak close to the magic number
$N=82$.

This might be at least partially explained by the fact that for this
simulation we have used an analytic continuous parameterization
(LMP(0)) for the EC rates washing out nuclear structure effects
observed in microscopic rate calculations \cite{LMP_ADNDT_2001}. Some
dependence on the different mass models might therefore be recovered
if microscopic rates consistent with the mass model were
used. Unfortunately, this can presently not be tested since, as we
show in the next section, the number of nuclear species present in the
tabulated microscopic rates is largely insufficient to cover the NSE
distribution and in particular the relevant nuclei for EC during
collapse. However, even if a final quantitative conclusion cannot be
drawn at present, it is clear from Fig.~\ref{fig:mass_model} that the
details of the mass model have much less influence on the dynamics of
core collapse than precise EC rates.

\section{Determination of the most relevant
  nuclei} \label{s:relevant_nuclei}

In the previous section we have shown that the most influential
microscopic ingredient entering a core-collapse simulation is the
expression of the individual EC rates, particularly their behavior at
low $Q$-values, which corresponds to very neutron-rich nuclei and
which is still largely unknown. The improved parameterization LMP(3),
providing a better fit to the microscopic calculations by
\citet{LMP_ADNDT_2001}, suggest a considerable average reduction of
the rates in the neutron-rich region with respect to the original
parameterization LMP(0). Still, the difference observed in the collapse
dynamics arises from the extrapolation of those fits to unknown
regions where no data nor microscopic calculations are available. It
is therefore clear that additional constraints are needed at low
$Q$-values before a parameterization can be considered as fully
reliable. For this reason, here we try to identify the most important
nuclei for the deleptonization process. Experimental and/or
microscopic calculations on these key nuclei could provide benchmarks
for future improved parameterizations.

\begin{figure}
  \resizebox{\columnwidth}{!}{\input{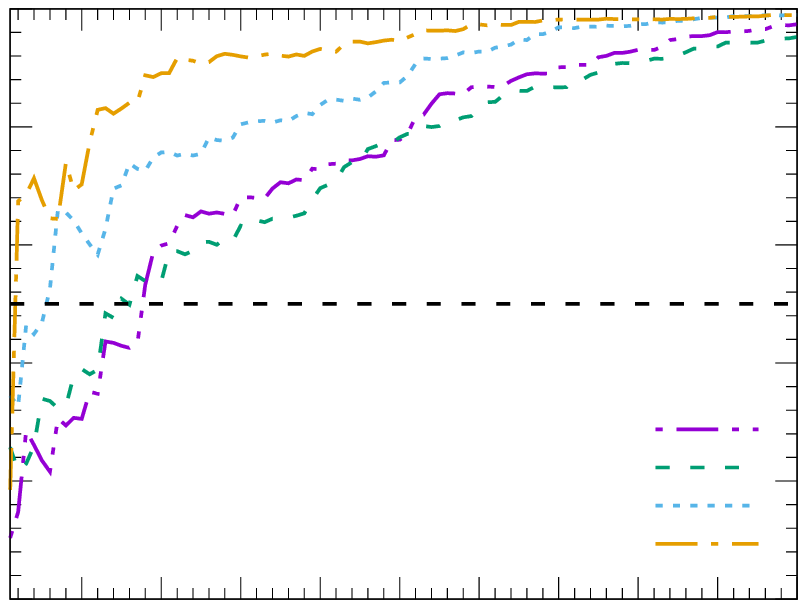}}   \resizebox{\columnwidth}{!}{\input{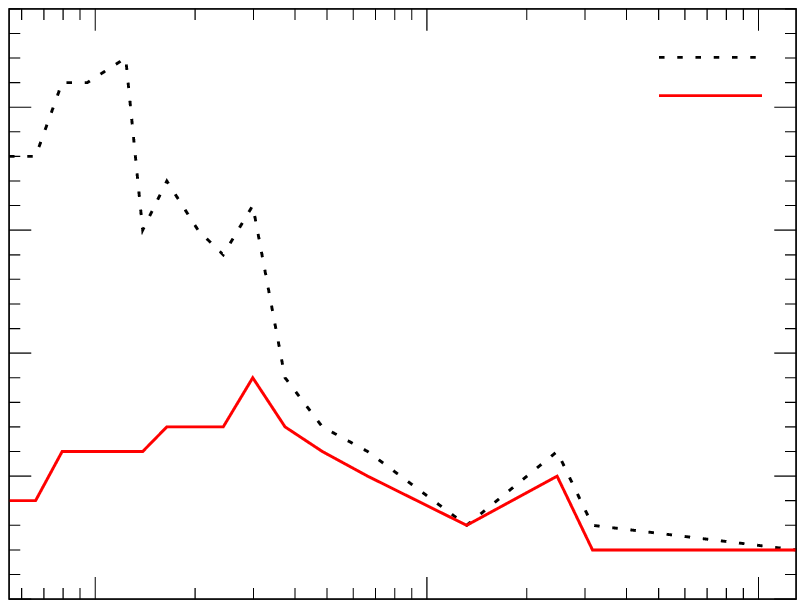}}
\caption{(color online) Upper panel: the instantaneous capture rate
  evaluated on a limited number $k$ of nuclei, normalized to the rate
  obtained taking the 200 most probable ones, as function of the size
  of the sample. Nuclei are ordered according to their abundance. The
  labels give the instantaneous rate on the whole distribution, lower
  rates corresponding to earlier times.
Lower panel: the number of nuclei accounting for 50\% of the total
instantaneous rate is plotted as a function of the EC rate (dashed
line). The number of species which are not included in the tabulated
microscopic EC rates~\cite{FFN_1982b, LMP_ADNDT_2001,
  Oda1994,Pruet2003} is plotted with a solid line.
  \label{fig:relevant_time}
}
\end{figure}

The simulations of this section were performed with the same settings
as in Sec.~\ref{ss:mass_model}, employing the HFB-24~\cite{gcp2013} mass model.

An estimation of the number of nuclear species that should be taken
into account to have a realistic core-collapse simulation can be
inferred from Fig.~\ref{fig:relevant_time}. In the upper part we
display the deleptonization rate obtained by considering only the $k$
most abundant nuclei in the NSE distribution, as a function of the
number $k$ of considered nuclei. The reason why we prefer to rank
nuclei according to their abundances rather than the more relevant
product between abundance and capture rate is that the latter quantity
is strongly affected by the assumed EC rate model. The different
curves are labelled by their instantaneous EC rate which is a
monotonically rising function of time, and the different curves
correspond thus to different times during collapse, before reaching
$\beta$-equilibrium. The rate is normalized to the value obtained by
summing the contribution of the 200 most abundant nuclei, considering
that $k=200$ is sufficient at all times to recover the total rate.

The SNA approximation, obtained considering $k=1$ in
Fig.~\ref{fig:relevant_time} (upper part), obviously leads to a
systematic underestimation of the rate by a factor between 2 and 10,
depending on the time. As observed in Fig.~\ref{fig:sna_nse}, this
underestimation leads to roughly 5\% overestimation of the electron
fraction at bounce, meaning that a larger pool of nuclei has to be
considered at each time step to have a complete picture of the
collapse.

To get an idea of how many different nuclei have to be considered, we
display in the lower part of Fig.~\ref{fig:relevant_time} the number of
nuclei responsible for half of the total EC rate as function of time
(labelled by the total instantaneous rate).
It shows that at
each time one half of the total rate is due to the capture on not more
than 20 nuclei, which reduce to less than 5 in the later stages before
$\beta$-equilibrium is reached and EC and its inverse $\beta$-decay become irrelevant.

We show in the same figure the number of those relevant nuclei for
which microscopic calculations~\cite{FFN_1982b, LMP_ADNDT_2001,
  Pruet2003, Oda1994} are not available (solid line). Evidently, for
many nuclei contributing in a dominant way to EC during collapse, no
microscopic rates exist. In particular in the later stages, most
relevant for the dynamics of the collapse, the rates on all those
nuclei have to be estimated by extrapolations.
\begin{figure}
  \includegraphics[width=\columnwidth]{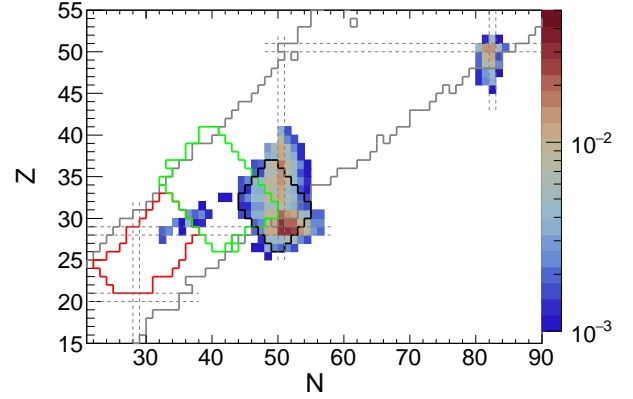}
    \caption{(color online) Time integrated relative deleptonization
      rate (color scale) associated to the different nuclear species
      identified by their proton $Z$ and neutron $N$ number. The black
      contour indicates the most relevant nuclei for EC identified by
      \textcite{Sullivan16} and \textcite{Titus2017}. The red, blue,
      and green contours indicate the nuclei for which microscopic
      rates are available from \textcite{LMP_ADNDT_2001},
      \textcite{Oda1994}, and \textcite{Pruet2003},
      respectively. Nuclei with experimentally known masses are
      situated between the grey lines.
  \label{fig:relevant_nuclei}
}
\end{figure}
Note, however, that the number of relevant unconstrained rates is
relatively limited. The corresponding isotopes are represented in
Fig.~\ref{fig:relevant_nuclei}, where the color scale represents the
relative contribution to the time integrated rate associated to the
different nuclei. The ensemble of nuclei represented in this figure
account for 89 \% of the total time integrated EC rate, and can
therefore be considered as the relevant pool of nuclei for the EC
process during infall. In the same figure, the grey lines delimit
the isotopic region where experimental values are known for the
nuclear masses, and the closed red/blue/green surfaces represents the
region where microscopic calculations are
available~\cite{LMP_ADNDT_2001,Pruet2003,Oda1994}. The nuclei relevant
for EC are essentially located close to the $N=50$ shell closure, in
good qualitative agreement with the results of the sensitivity study
by \textcite{Sullivan16} and \textcite{Titus2017} (shown by the black
contour).

It should be stressed that for most of 130 different nuclei
represented in Fig.~\ref{fig:relevant_nuclei} no microscopic
calculation exists. Dedicated microscopic calculations for all these
nuclei represent still an enormous challenge, but it is interesting to
observe that these nuclei are concentrated on a relatively reduced
zone of the nuclear chart. This means that some extra experimental
and/or theoretical information on weak processes even on a few of the
most important ones would greatly help to constrain analytic forms
such as LMP(3), for a systematic application to the whole nuclear
pool.

One remark of caution is needed at his point. For our calculations
here, we have used the LMP(0) parameterization for all nuclei. This is
at variance with \citet{Sullivan16}, where the LMP(0) was used only
for nuclei that are not contained in any of the three tables of
Refs.~\cite{Oda1994, LMP_ADNDT_2001, Pruet2003}. The exact list of
relevant nuclei for EC might thus be slightly different from those
shown in Fig.~\ref{fig:relevant_nuclei} if the microscopic rates were
used where they are known. We expect this effect, however, to be
small since most of the relevant nuclei lie outside the range of the
tables with microscopic rates. Please note that the absolute rates
reported in Fig.~\ref{fig:relevant_nuclei} are anyway model
dependent. For instance, if we had employed the LMP(3) parameterization
instead of LMP(0) they would clearly be reduced.

\section{Conclusions}\label{s:conc}
Within this study we have performed simulations of the pre-bounce
evolution of core-collapse supernovae investigating the effect of
improved EC rates on nuclei developed in \textcite{Raduta2017}. As
pointed out already by \citet{Langanke_PRL_2003}, Bruenn's EC rates
commonly used break down as soon as neutron-rich nuclei become
populated abundantly in the later stages of the pre-bounce evolution.
\ar{Although some attempts have been made to extent LSSM calculations
  to heavier and more neutron rich nuclei using hybrid
  approaches~\cite{Juodagalvis_NPA_2010,Dean_PRC_1998},
  microscopically calculated rates for these nuclei are still not the
  most convenient to be directly applied in large CCSN simulations.}
Therefore, for the most exotic and neutron-rich nuclei EC rates
\ar{are} extrapolated~\cite{Langanke_PRL_2003} and are thus
subject to large uncertainties. \textcite{Sullivan16} have therefore
performed a systematic sensitivity study, where individual EC rates
were thereby globally scaled by arbitrary factors ranging from 2 to 10
with respect to the fiducial values showing that the uncertainties on
nuclear EC rates have a stronger influence on pre-bounce evolution
than other inputs, such as progenitor model or EoS.

Here, we confirm qualitatively the findings as well of the pioneering
work of \citet{Langanke_PRL_2003, Hix2003} as that of \citet{Sullivan16}
and the subsequent studies of~\cite{Titus2017}. Electron captures
occur predominantly on neutron-rich nuclei during the last stages of
pre-bounce and enhanced captures reduce central $Y_e$ at bounce with
differences of up to 30\% between the different prescriptions for EC
on heavy neutron-rich nuclei. Lower $Y_e$ at bounce leads to smaller
inner core mass and slower shock propagation after bounce. The effect
of different EC rates is clearly predominant with respect to the EoS,
the nuclear mass model, or the progenitor model. However, the
improved parameterization LMP(3) motivated by nuclear physics
considerations clearly points to a reduction of EC on neutron-rich
nuclei with respect to the work by \textcite{Langanke_PRL_2003}.
Although the impact of EC on neutron-rich nuclei is thus attenuated,
we emphasize that still, the results are considerably different from
those employing Bruenn's rates and it is important to include EC on
those nuclei.

The important sensitivity to the different prescriptions for EC rates
clearly indicates the importance of clarifying the rates on those
nuclei, either by theoretical calculations or experiments. Indeed,
the parameterization LMP(3) is certainly improved by nuclear physics
considerations with respect to the basic extrapolation proposed in
LMP(0), but still it is just an improved fit to complete LSSM
calculations, and the differences with respect to the simpler LMP(0)
prescription arise from the extrapolation of the fit to the low
$Q$-value region where no microscopic calculations are available.
We have therefore provided a list of the most relevant nuclei
accounting for the 89 \% of the total time integrated deleptonization
rate. Although the details of this list are certainly model dependent
(EoS, mass model, progenitor, EC rate parameterization, \dots), there
is a large overlap with the 74 most important nuclei for EC identified
in Refs.~\cite{Sullivan16,Titus2017} with different settings, such that the
identification is robust.

\ar{Our approach on the computation of EC rates in CCSN simulatons
  presented here possesses the advantages of being numerically
  efficient and applicable to any EoS. We plan to make public in the
  near future these rates as well as neutrino-nucleus scattering
  opacities.}

\begin{acknowledgments}
  We would like to thank J. Pons for providing us with the original
  version of the ACCEPT code. The research leading to these results
  has received funding from the PICS07889; it was also partially
  supported by the PHAROS European Science and Technology (COST)
  Action CA16214 and the Observatoire de Paris through the PTV
  programme and the action fédératrice ``PhyFog''.
\end{acknowledgments}
\bibliography{biblio}

\begin{thebibliography}{73}%
\makeatletter
\providecommand \@ifxundefined [1]{%
 \@ifx{#1\undefined}
}%
\providecommand \@ifnum [1]{%
 \ifnum #1\expandafter \@firstoftwo
 \else \expandafter \@secondoftwo
 \fi
}%
\providecommand \@ifx [1]{%
 \ifx #1\expandafter \@firstoftwo
 \else \expandafter \@secondoftwo
 \fi
}%
\providecommand \natexlab [1]{#1}%
\providecommand \enquote  [1]{``#1''}%
\providecommand \bibnamefont  [1]{#1}%
\providecommand \bibfnamefont [1]{#1}%
\providecommand \citenamefont [1]{#1}%
\providecommand \href@noop [0]{\@secondoftwo}%
\providecommand \href [0]{\begingroup \@sanitize@url \@href}%
\providecommand \@href[1]{\@@startlink{#1}\@@href}%
\providecommand \@@href[1]{\endgroup#1\@@endlink}%
\providecommand \@sanitize@url [0]{\catcode `\\12\catcode `\$12\catcode
  `\&12\catcode `\#12\catcode `\^12\catcode `\_12\catcode `\%12\relax}%
\providecommand \@@startlink[1]{}%
\providecommand \@@endlink[0]{}%
\providecommand \url  [0]{\begingroup\@sanitize@url \@url }%
\providecommand \@url [1]{\endgroup\@href {#1}{\urlprefix }}%
\providecommand \urlprefix  [0]{URL }%
\providecommand \Eprint [0]{\href }%
\providecommand \doibase [0]{http://dx.doi.org/}%
\providecommand \selectlanguage [0]{\@gobble}%
\providecommand \bibinfo  [0]{\@secondoftwo}%
\providecommand \bibfield  [0]{\@secondoftwo}%
\providecommand \translation [1]{[#1]}%
\providecommand \BibitemOpen [0]{}%
\providecommand \bibitemStop [0]{}%
\providecommand \bibitemNoStop [0]{.\EOS\space}%
\providecommand \EOS [0]{\spacefactor3000\relax}%
\providecommand \BibitemShut  [1]{\csname bibitem#1\endcsname}%
\let\auto@bib@innerbib\@empty
\bibitem [{\citenamefont {Janka}(2012)}]{Janka2012a}%
  \BibitemOpen
  \bibfield  {author} {\bibinfo {author} {\bibfnamefont {H.-T.}\ \bibnamefont
  {Janka}},\ }\href {\doibase 10.1146/annurev-nucl-102711-094901} {\bibfield
  {journal} {\bibinfo  {journal} {Ann. Rev. Nucl. Part. Sci.}\ }\textbf
  {\bibinfo {volume} {62}},\ \bibinfo {pages} {407} (\bibinfo {year} {2012})},\
  \Eprint {http://arxiv.org/abs/1206.2503} {arXiv:1206.2503 [astro-ph.SR]}
  \BibitemShut {NoStop}%
\bibitem [{\citenamefont {{Janka}}(2017)}]{Janka2017b}%
  \BibitemOpen
  \bibfield  {author} {\bibinfo {author} {\bibfnamefont {H.-T.}\ \bibnamefont
  {{Janka}}},\ }\enquote {\bibinfo {title} {{Neutrino-Driven Explosions}},}\
  in\ \href {\doibase 10.1007/978-3-319-21846-5-109} {\emph {\bibinfo
  {booktitle} {Handbook of Supernovae, ISBN 978-3-319-21845-8. Springer
  International Publishing AG, 2017, p. 1095}}}\ (\bibinfo {year} {2017})\ p.\
  \bibinfo {pages} {1095}\BibitemShut {NoStop}%
\bibitem [{\citenamefont {Bethe}\ \emph {et~al.}(1979)\citenamefont {Bethe},
  \citenamefont {Brown}, \citenamefont {Applegate},\ and\ \citenamefont
  {Lattimer}}]{Bethe_NPA_1979}%
  \BibitemOpen
  \bibfield  {author} {\bibinfo {author} {\bibfnamefont {H.}~\bibnamefont
  {Bethe}}, \bibinfo {author} {\bibfnamefont {G.}~\bibnamefont {Brown}},
  \bibinfo {author} {\bibfnamefont {J.}~\bibnamefont {Applegate}}, \ and\
  \bibinfo {author} {\bibfnamefont {J.}~\bibnamefont {Lattimer}},\ }\href
  {\doibase https://doi.org/10.1016/0375-9474(79)90596-7} {\bibfield  {journal}
  {\bibinfo  {journal} {Nuclear Physics A}\ }\textbf {\bibinfo {volume}
  {324}},\ \bibinfo {pages} {487 } (\bibinfo {year} {1979})}\BibitemShut
  {NoStop}%
\bibitem [{\citenamefont {{Aufderheide}}\ \emph {et~al.}(1994)\citenamefont
  {{Aufderheide}}, \citenamefont {{Fushiki}}, \citenamefont {{Woosley}},\ and\
  \citenamefont {{Hartmann}}}]{Aufderheide94}%
  \BibitemOpen
  \bibfield  {author} {\bibinfo {author} {\bibfnamefont {M.~B.}\ \bibnamefont
  {{Aufderheide}}}, \bibinfo {author} {\bibfnamefont {I.}~\bibnamefont
  {{Fushiki}}}, \bibinfo {author} {\bibfnamefont {S.~E.}\ \bibnamefont
  {{Woosley}}}, \ and\ \bibinfo {author} {\bibfnamefont {D.~H.}\ \bibnamefont
  {{Hartmann}}},\ }\href {\doibase 10.1086/191942} {\bibfield  {journal}
  {\bibinfo  {journal} {Astrophys. J. Suppl.}\ }\textbf {\bibinfo {volume}
  {91}},\ \bibinfo {pages} {389} (\bibinfo {year} {1994})}\BibitemShut
  {NoStop}%
\bibitem [{\citenamefont {Heger}\ \emph
  {et~al.}(2001{\natexlab{a}})\citenamefont {Heger}, \citenamefont {Langanke},
  \citenamefont {Martinez-Pinedo},\ and\ \citenamefont {Woosley}}]{Heger2001a}%
  \BibitemOpen
  \bibfield  {author} {\bibinfo {author} {\bibfnamefont {A.}~\bibnamefont
  {Heger}}, \bibinfo {author} {\bibfnamefont {K.}~\bibnamefont {Langanke}},
  \bibinfo {author} {\bibfnamefont {G.}~\bibnamefont {Martinez-Pinedo}}, \ and\
  \bibinfo {author} {\bibfnamefont {S.~E.}\ \bibnamefont {Woosley}},\ }\href
  {\doibase 10.1103/PhysRevLett.86.1678} {\bibfield  {journal} {\bibinfo
  {journal} {Phys. Rev. Lett.}\ }\textbf {\bibinfo {volume} {86}},\ \bibinfo
  {pages} {1678} (\bibinfo {year} {2001}{\natexlab{a}})}\BibitemShut {NoStop}%
\bibitem [{\citenamefont {Heger}\ \emph
  {et~al.}(2001{\natexlab{b}})\citenamefont {Heger}, \citenamefont {Woosley},
  \citenamefont {Martinez-Pinedo},\ and\ \citenamefont
  {Langanke}}]{Heger2001b}%
  \BibitemOpen
  \bibfield  {author} {\bibinfo {author} {\bibfnamefont {A.}~\bibnamefont
  {Heger}}, \bibinfo {author} {\bibfnamefont {S.~E.}\ \bibnamefont {Woosley}},
  \bibinfo {author} {\bibfnamefont {G.}~\bibnamefont {Martinez-Pinedo}}, \ and\
  \bibinfo {author} {\bibfnamefont {K.}~\bibnamefont {Langanke}},\ }\href
  {\doibase 10.1086/324092} {\bibfield  {journal} {\bibinfo  {journal}
  {Astrophys. J.}\ }\textbf {\bibinfo {volume} {560}},\ \bibinfo {pages} {307}
  (\bibinfo {year} {2001}{\natexlab{b}})}\BibitemShut {NoStop}%
\bibitem [{\citenamefont {Hix}\ \emph {et~al.}(2003)\citenamefont {Hix},
  \citenamefont {Messer}, \citenamefont {Mezzacappa}, \citenamefont
  {Liebend{\"o}rfer}, \citenamefont {Sampaio}, \citenamefont {Langanke},
  \citenamefont {Dean},\ and\ \citenamefont {Martinez-Pinedo}}]{Hix2003}%
  \BibitemOpen
  \bibfield  {author} {\bibinfo {author} {\bibfnamefont {W.~R.}\ \bibnamefont
  {Hix}}, \bibinfo {author} {\bibfnamefont {O.~E.~B.}\ \bibnamefont {Messer}},
  \bibinfo {author} {\bibfnamefont {A.}~\bibnamefont {Mezzacappa}}, \bibinfo
  {author} {\bibfnamefont {M.}~\bibnamefont {Liebend{\"o}rfer}}, \bibinfo
  {author} {\bibfnamefont {J.}~\bibnamefont {Sampaio}}, \bibinfo {author}
  {\bibfnamefont {K.}~\bibnamefont {Langanke}}, \bibinfo {author}
  {\bibfnamefont {D.~J.}\ \bibnamefont {Dean}}, \ and\ \bibinfo {author}
  {\bibfnamefont {G.}~\bibnamefont {Martinez-Pinedo}},\ }\href {\doibase
  10.1103/PhysRevLett.91.201102} {\bibfield  {journal} {\bibinfo  {journal}
  {Phys. Rev. Lett.}\ }\textbf {\bibinfo {volume} {91}},\ \bibinfo {pages}
  {201102} (\bibinfo {year} {2003})}\BibitemShut {NoStop}%
\bibitem [{\citenamefont {Janka}\ \emph {et~al.}(2007)\citenamefont {Janka},
  \citenamefont {Langanke}, \citenamefont {Marek}, \citenamefont
  {Martinez-Pinedo},\ and\ \citenamefont {Mueller}}]{Janka2007}%
  \BibitemOpen
  \bibfield  {author} {\bibinfo {author} {\bibfnamefont {H.-T.}\ \bibnamefont
  {Janka}}, \bibinfo {author} {\bibfnamefont {K.}~\bibnamefont {Langanke}},
  \bibinfo {author} {\bibfnamefont {A.}~\bibnamefont {Marek}}, \bibinfo
  {author} {\bibfnamefont {G.}~\bibnamefont {Martinez-Pinedo}}, \ and\ \bibinfo
  {author} {\bibfnamefont {B.}~\bibnamefont {Mueller}},\ }\href {\doibase
  10.1016/j.physrep.2007.02.002} {\bibfield  {journal} {\bibinfo  {journal}
  {Phys. Rept.}\ }\textbf {\bibinfo {volume} {442}},\ \bibinfo {pages} {38}
  (\bibinfo {year} {2007})}\BibitemShut {NoStop}%
\bibitem [{\citenamefont {{DUNE Collaboration}}(2018)}]{DUNE2018}%
  \BibitemOpen
  \bibfield  {author} {\bibinfo {author} {\bibnamefont {{DUNE
  Collaboration}}},\ }\href@noop {} {\bibfield  {journal} {\bibinfo  {journal}
  {arXiv e-prints}\ ,\ \bibinfo {eid} {arXiv:1807.10334}} (\bibinfo {year}
  {2018})},\ \Eprint {http://arxiv.org/abs/1807.10334} {arXiv:1807.10334
  [physics.ins-det]} \BibitemShut {NoStop}%
\bibitem [{\citenamefont {Kato}\ \emph {et~al.}(2017)\citenamefont {Kato},
  \citenamefont {Nagakura}, \citenamefont {Furusawa}, \citenamefont
  {Takahashi}, \citenamefont {Umeda}, \citenamefont {Yoshida}, \citenamefont
  {Ishidoshiro},\ and\ \citenamefont {Yamada}}]{Kato2017}%
  \BibitemOpen
  \bibfield  {author} {\bibinfo {author} {\bibfnamefont {C.}~\bibnamefont
  {Kato}}, \bibinfo {author} {\bibfnamefont {H.}~\bibnamefont {Nagakura}},
  \bibinfo {author} {\bibfnamefont {S.}~\bibnamefont {Furusawa}}, \bibinfo
  {author} {\bibfnamefont {K.}~\bibnamefont {Takahashi}}, \bibinfo {author}
  {\bibfnamefont {H.}~\bibnamefont {Umeda}}, \bibinfo {author} {\bibfnamefont
  {T.}~\bibnamefont {Yoshida}}, \bibinfo {author} {\bibfnamefont
  {K.}~\bibnamefont {Ishidoshiro}}, \ and\ \bibinfo {author} {\bibfnamefont
  {S.}~\bibnamefont {Yamada}},\ }\href {\doibase 10.3847/1538-4357/aa8b72}
  {\bibfield  {journal} {\bibinfo  {journal} {Astrophys. J.}\ }\textbf
  {\bibinfo {volume} {848}},\ \bibinfo {pages} {48} (\bibinfo {year}
  {2017})}\BibitemShut {NoStop}%
\bibitem [{\citenamefont {Juodagalvis}\ \emph {et~al.}(2010)\citenamefont
  {Juodagalvis}, \citenamefont {Langanke}, \citenamefont {Hix}, \citenamefont
  {Martínez-Pinedo},\ and\ \citenamefont {Sampaio}}]{Juodagalvis_NPA_2010}%
  \BibitemOpen
  \bibfield  {author} {\bibinfo {author} {\bibfnamefont {A.}~\bibnamefont
  {Juodagalvis}}, \bibinfo {author} {\bibfnamefont {K.}~\bibnamefont
  {Langanke}}, \bibinfo {author} {\bibfnamefont {W.}~\bibnamefont {Hix}},
  \bibinfo {author} {\bibfnamefont {G.}~\bibnamefont {Martínez-Pinedo}}, \
  and\ \bibinfo {author} {\bibfnamefont {J.}~\bibnamefont {Sampaio}},\ }\href
  {\doibase https://doi.org/10.1016/j.nuclphysa.2010.09.012} {\bibfield
  {journal} {\bibinfo  {journal} {Nuclear Physics A}\ }\textbf {\bibinfo
  {volume} {848}},\ \bibinfo {pages} {454 } (\bibinfo {year}
  {2010})}\BibitemShut {NoStop}%
\bibitem [{\citenamefont {Sullivan}\ \emph {et~al.}(2016)\citenamefont
  {Sullivan}, \citenamefont {O'Connor}, \citenamefont {Zegers}, \citenamefont
  {Grubb},\ and\ \citenamefont {Austin}}]{Sullivan16}%
  \BibitemOpen
  \bibfield  {author} {\bibinfo {author} {\bibfnamefont {C.}~\bibnamefont
  {Sullivan}}, \bibinfo {author} {\bibfnamefont {E.}~\bibnamefont {O'Connor}},
  \bibinfo {author} {\bibfnamefont {R.~G.~T.}\ \bibnamefont {Zegers}}, \bibinfo
  {author} {\bibfnamefont {T.}~\bibnamefont {Grubb}}, \ and\ \bibinfo {author}
  {\bibfnamefont {S.~M.}\ \bibnamefont {Austin}},\ }\href {\doibase
  10.3847/0004-637X/816/1/44} {\bibfield  {journal} {\bibinfo  {journal}
  {Astrophys. J.}\ }\textbf {\bibinfo {volume} {816}},\ \bibinfo {pages} {44}
  (\bibinfo {year} {2016})}\BibitemShut {NoStop}%
\bibitem [{\citenamefont {Raduta}\ \emph {et~al.}(2016)\citenamefont {Raduta},
  \citenamefont {Gulminelli},\ and\ \citenamefont {Oertel}}]{Raduta2016}%
  \BibitemOpen
  \bibfield  {author} {\bibinfo {author} {\bibfnamefont {A.~R.}\ \bibnamefont
  {Raduta}}, \bibinfo {author} {\bibfnamefont {F.}~\bibnamefont {Gulminelli}},
  \ and\ \bibinfo {author} {\bibfnamefont {M.}~\bibnamefont {Oertel}},\ }\href
  {\doibase 10.1103/PhysRevC.93.025803} {\bibfield  {journal} {\bibinfo
  {journal} {Phys. Rev.}\ }\textbf {\bibinfo {volume} {C93}},\ \bibinfo {pages}
  {025803} (\bibinfo {year} {2016})}\BibitemShut {NoStop}%
\bibitem [{\citenamefont {Furusawa}\ \emph {et~al.}(2017)\citenamefont
  {Furusawa}, \citenamefont {Nagakura}, \citenamefont {Sumiyoshi},
  \citenamefont {Kato},\ and\ \citenamefont {Yamada}}]{Furusawa2017}%
  \BibitemOpen
  \bibfield  {author} {\bibinfo {author} {\bibfnamefont {S.}~\bibnamefont
  {Furusawa}}, \bibinfo {author} {\bibfnamefont {H.}~\bibnamefont {Nagakura}},
  \bibinfo {author} {\bibfnamefont {K.}~\bibnamefont {Sumiyoshi}}, \bibinfo
  {author} {\bibfnamefont {C.}~\bibnamefont {Kato}}, \ and\ \bibinfo {author}
  {\bibfnamefont {S.}~\bibnamefont {Yamada}},\ }\href {\doibase
  10.1103/PhysRevC.95.025809} {\bibfield  {journal} {\bibinfo  {journal} {Phys.
  Rev.}\ }\textbf {\bibinfo {volume} {C95}},\ \bibinfo {pages} {025809}
  (\bibinfo {year} {2017})}\BibitemShut {NoStop}%
\bibitem [{\citenamefont {Raduta}\ \emph {et~al.}(2017)\citenamefont {Raduta},
  \citenamefont {Gulminelli},\ and\ \citenamefont {Oertel}}]{Raduta2017}%
  \BibitemOpen
  \bibfield  {author} {\bibinfo {author} {\bibfnamefont {A.~R.}\ \bibnamefont
  {Raduta}}, \bibinfo {author} {\bibfnamefont {F.}~\bibnamefont {Gulminelli}},
  \ and\ \bibinfo {author} {\bibfnamefont {M.}~\bibnamefont {Oertel}},\ }\href
  {\doibase 10.1103/PhysRevC.95.025805} {\bibfield  {journal} {\bibinfo
  {journal} {Phys. Rev.}\ }\textbf {\bibinfo {volume} {C95}},\ \bibinfo {pages}
  {025805} (\bibinfo {year} {2017})}\BibitemShut {NoStop}%
\bibitem [{\citenamefont {Yudin}\ \emph {et~al.}(2018)\citenamefont {Yudin},
  \citenamefont {Hempel}, \citenamefont {Blinnikov}, \citenamefont
  {Nadyozhin},\ and\ \citenamefont {Panov}}]{Yudin2018}%
  \BibitemOpen
  \bibfield  {author} {\bibinfo {author} {\bibfnamefont {A.}~\bibnamefont
  {Yudin}}, \bibinfo {author} {\bibfnamefont {M.}~\bibnamefont {Hempel}},
  \bibinfo {author} {\bibfnamefont {S.}~\bibnamefont {Blinnikov}}, \bibinfo
  {author} {\bibfnamefont {D.}~\bibnamefont {Nadyozhin}}, \ and\ \bibinfo
  {author} {\bibfnamefont {I.}~\bibnamefont {Panov}},\ }\href {\doibase
  10.1093/mnras/sty3468} {\  (\bibinfo {year} {2018}),\
  10.1093/mnras/sty3468},\ \Eprint {http://arxiv.org/abs/1812.09494}
  {arXiv:1812.09494 [nucl-th]} \BibitemShut {NoStop}%
\bibitem [{\citenamefont {Nagakura}\ \emph {et~al.}(2019)\citenamefont
  {Nagakura}, \citenamefont {Furusawa}, \citenamefont {Togashi}, \citenamefont
  {Richers}, \citenamefont {Sumiyoshi},\ and\ \citenamefont
  {Yamada}}]{Nagakura2019}%
  \BibitemOpen
  \bibfield  {author} {\bibinfo {author} {\bibfnamefont {H.}~\bibnamefont
  {Nagakura}}, \bibinfo {author} {\bibfnamefont {S.}~\bibnamefont {Furusawa}},
  \bibinfo {author} {\bibfnamefont {H.}~\bibnamefont {Togashi}}, \bibinfo
  {author} {\bibfnamefont {S.}~\bibnamefont {Richers}}, \bibinfo {author}
  {\bibfnamefont {K.}~\bibnamefont {Sumiyoshi}}, \ and\ \bibinfo {author}
  {\bibfnamefont {S.}~\bibnamefont {Yamada}},\ }\href {\doibase
  10.3847/1538-4365/aafac9} {\bibfield  {journal} {\bibinfo  {journal}
  {Astrophys. J. Suppl.}\ }\textbf {\bibinfo {volume} {240}},\ \bibinfo {pages}
  {38} (\bibinfo {year} {2019})}\BibitemShut {NoStop}%
\bibitem [{\citenamefont {Titus}\ \emph {et~al.}(2018)\citenamefont {Titus},
  \citenamefont {Sullivan}, \citenamefont {Zegers}, \citenamefont {Brown},\
  and\ \citenamefont {Gao}}]{Titus2017}%
  \BibitemOpen
  \bibfield  {author} {\bibinfo {author} {\bibfnamefont {R.}~\bibnamefont
  {Titus}}, \bibinfo {author} {\bibfnamefont {C.}~\bibnamefont {Sullivan}},
  \bibinfo {author} {\bibfnamefont {R.~G.~T.}\ \bibnamefont {Zegers}}, \bibinfo
  {author} {\bibfnamefont {B.~A.}\ \bibnamefont {Brown}}, \ and\ \bibinfo
  {author} {\bibfnamefont {B.}~\bibnamefont {Gao}},\ }\href {\doibase
  10.1088/1361-6471/aa98c1} {\bibfield  {journal} {\bibinfo  {journal} {J.
  Phys.}\ }\textbf {\bibinfo {volume} {G45}},\ \bibinfo {pages} {014004}
  (\bibinfo {year} {2018})}\BibitemShut {NoStop}%
\bibitem [{\citenamefont {{Titus}}\ \emph {et~al.}(2019)\citenamefont
  {{Titus}}, \citenamefont {{Ney}}, \citenamefont {{Zegers}}, \citenamefont
  {{Bazin}}, \citenamefont {{Belarge}}, \citenamefont {{Bender}}, \citenamefont
  {{Brown}}, \citenamefont {{Campbell}}, \citenamefont {{Elman}}, \citenamefont
  {{Engel}}, \citenamefont {{Gade}}, \citenamefont {{Gao}}, \citenamefont
  {{Kwan}}, \citenamefont {{Lipschutz}}, \citenamefont {{Longfellow}},
  \citenamefont {{Lunderberg}}, \citenamefont {{Mijatovic}}, \citenamefont
  {{Noji}}, \citenamefont {{Pereira}}, \citenamefont {{Schmitt}}, \citenamefont
  {{Sullivan}}, \citenamefont {{Weisshaar}},\ and\ \citenamefont
  {{Zamora}}}]{Titus2019}%
  \BibitemOpen
  \bibfield  {author} {\bibinfo {author} {\bibfnamefont {R.}~\bibnamefont
  {{Titus}}}, \bibinfo {author} {\bibfnamefont {E.~M.}\ \bibnamefont {{Ney}}},
  \bibinfo {author} {\bibfnamefont {R.~G.~T.}\ \bibnamefont {{Zegers}}},
  \bibinfo {author} {\bibfnamefont {D.}~\bibnamefont {{Bazin}}}, \bibinfo
  {author} {\bibfnamefont {J.}~\bibnamefont {{Belarge}}}, \bibinfo {author}
  {\bibfnamefont {P.~C.}\ \bibnamefont {{Bender}}}, \bibinfo {author}
  {\bibfnamefont {B.~A.}\ \bibnamefont {{Brown}}}, \bibinfo {author}
  {\bibfnamefont {C.~M.}\ \bibnamefont {{Campbell}}}, \bibinfo {author}
  {\bibfnamefont {B.}~\bibnamefont {{Elman}}}, \bibinfo {author} {\bibfnamefont
  {J.}~\bibnamefont {{Engel}}}, \bibinfo {author} {\bibfnamefont
  {A.}~\bibnamefont {{Gade}}}, \bibinfo {author} {\bibfnamefont
  {B.}~\bibnamefont {{Gao}}}, \bibinfo {author} {\bibfnamefont
  {E.}~\bibnamefont {{Kwan}}}, \bibinfo {author} {\bibfnamefont
  {S.}~\bibnamefont {{Lipschutz}}}, \bibinfo {author} {\bibfnamefont
  {B.}~\bibnamefont {{Longfellow}}}, \bibinfo {author} {\bibfnamefont
  {E.}~\bibnamefont {{Lunderberg}}}, \bibinfo {author} {\bibfnamefont
  {T.}~\bibnamefont {{Mijatovic}}}, \bibinfo {author} {\bibfnamefont
  {S.}~\bibnamefont {{Noji}}}, \bibinfo {author} {\bibfnamefont
  {J.}~\bibnamefont {{Pereira}}}, \bibinfo {author} {\bibfnamefont
  {J.}~\bibnamefont {{Schmitt}}}, \bibinfo {author} {\bibfnamefont
  {C.}~\bibnamefont {{Sullivan}}}, \bibinfo {author} {\bibfnamefont
  {D.}~\bibnamefont {{Weisshaar}}}, \ and\ \bibinfo {author} {\bibfnamefont
  {J.~C.}\ \bibnamefont {{Zamora}}},\ }\href@noop {} {\bibfield  {journal}
  {\bibinfo  {journal} {arXiv e-prints}\ ,\ \bibinfo {eid} {arXiv:1908.03985}}
  (\bibinfo {year} {2019})},\ \Eprint {http://arxiv.org/abs/1908.03985}
  {arXiv:1908.03985 [nucl-ex]} \BibitemShut {NoStop}%
\bibitem [{\citenamefont {{Fuller}}\ \emph
  {et~al.}(1982{\natexlab{a}})\citenamefont {{Fuller}}, \citenamefont
  {{Fowler}},\ and\ \citenamefont {{Newman}}}]{FFN_1982b}%
  \BibitemOpen
  \bibfield  {author} {\bibinfo {author} {\bibfnamefont {G.~M.}\ \bibnamefont
  {{Fuller}}}, \bibinfo {author} {\bibfnamefont {W.~A.}\ \bibnamefont
  {{Fowler}}}, \ and\ \bibinfo {author} {\bibfnamefont {M.~J.}\ \bibnamefont
  {{Newman}}},\ }\href {\doibase 10.1086/190779} {\bibfield  {journal}
  {\bibinfo  {journal} {Astrophys. J. Suppl.}\ }\textbf {\bibinfo {volume}
  {48}},\ \bibinfo {pages} {279} (\bibinfo {year}
  {1982}{\natexlab{a}})}\BibitemShut {NoStop}%
\bibitem [{\citenamefont {Langanke}\ and\ \citenamefont
  {Martinez-Pinedo}(2001)}]{LMP_ADNDT_2001}%
  \BibitemOpen
  \bibfield  {author} {\bibinfo {author} {\bibfnamefont {K.}~\bibnamefont
  {Langanke}}\ and\ \bibinfo {author} {\bibfnamefont {G.}~\bibnamefont
  {Martinez-Pinedo}},\ }\href {\doibase 10.1006/adnd.2001.0865} {\bibfield
  {journal} {\bibinfo  {journal} {Atom. Data Nucl. Data Tabl.}\ }\textbf
  {\bibinfo {volume} {79}},\ \bibinfo {pages} {1} (\bibinfo {year}
  {2001})}\BibitemShut {NoStop}%
\bibitem [{\citenamefont {Langanke}\ and\ \citenamefont
  {Martinez-Pinedo}(2003)}]{Langanke2002}%
  \BibitemOpen
  \bibfield  {author} {\bibinfo {author} {\bibfnamefont {K.}~\bibnamefont
  {Langanke}}\ and\ \bibinfo {author} {\bibfnamefont {G.}~\bibnamefont
  {Martinez-Pinedo}},\ }\href {\doibase 10.1103/RevModPhys.75.819} {\bibfield
  {journal} {\bibinfo  {journal} {Rev. Mod. Phys.}\ }\textbf {\bibinfo {volume}
  {75}},\ \bibinfo {pages} {819} (\bibinfo {year} {2003})}\BibitemShut
  {NoStop}%
\bibitem [{\citenamefont {Oda}\ \emph {et~al.}(1994)\citenamefont {Oda},
  \citenamefont {Hino}, \citenamefont {Muto}, \citenamefont {Takahara},\ and\
  \citenamefont {Sato}}]{Oda1994}%
  \BibitemOpen
  \bibfield  {author} {\bibinfo {author} {\bibfnamefont {T.}~\bibnamefont
  {Oda}}, \bibinfo {author} {\bibfnamefont {M.}~\bibnamefont {Hino}}, \bibinfo
  {author} {\bibfnamefont {K.}~\bibnamefont {Muto}}, \bibinfo {author}
  {\bibfnamefont {M.}~\bibnamefont {Takahara}}, \ and\ \bibinfo {author}
  {\bibfnamefont {K.}~\bibnamefont {Sato}},\ }\href {\doibase
  10.1006/adnd.1994.1007} {\bibfield  {journal} {\bibinfo  {journal} {Atom.
  Data Nucl. Data Tabl.}\ }\textbf {\bibinfo {volume} {56}},\ \bibinfo {pages}
  {231} (\bibinfo {year} {1994})}\BibitemShut {NoStop}%
\bibitem [{\citenamefont {Pruet}\ and\ \citenamefont
  {Fuller}(2003)}]{Pruet2003}%
  \BibitemOpen
  \bibfield  {author} {\bibinfo {author} {\bibfnamefont {J.}~\bibnamefont
  {Pruet}}\ and\ \bibinfo {author} {\bibfnamefont {G.~M.}\ \bibnamefont
  {Fuller}},\ }\href {\doibase 10.1086/376753} {\bibfield  {journal} {\bibinfo
  {journal} {Astrophys. J. Suppl.}\ }\textbf {\bibinfo {volume} {149}},\
  \bibinfo {pages} {189} (\bibinfo {year} {2003})}\BibitemShut {NoStop}%
\bibitem [{\citenamefont {Langanke}\ \emph {et~al.}(2003)\citenamefont
  {Langanke}, \citenamefont {Martinez-Pinedo}, \citenamefont {Sampaio},
  \citenamefont {Dean}, \citenamefont {Hix}, \citenamefont {Messer},
  \citenamefont {Mezzacappa}, \citenamefont {Liebend{\"o}rfer}, \citenamefont
  {Janka},\ and\ \citenamefont {Rampp}}]{Langanke_PRL_2003}%
  \BibitemOpen
  \bibfield  {author} {\bibinfo {author} {\bibfnamefont {K.}~\bibnamefont
  {Langanke}}, \bibinfo {author} {\bibfnamefont {G.}~\bibnamefont
  {Martinez-Pinedo}}, \bibinfo {author} {\bibfnamefont {J.~M.}\ \bibnamefont
  {Sampaio}}, \bibinfo {author} {\bibfnamefont {D.~J.}\ \bibnamefont {Dean}},
  \bibinfo {author} {\bibfnamefont {W.~R.}\ \bibnamefont {Hix}}, \bibinfo
  {author} {\bibfnamefont {O.~E.~B.}\ \bibnamefont {Messer}}, \bibinfo {author}
  {\bibfnamefont {A.}~\bibnamefont {Mezzacappa}}, \bibinfo {author}
  {\bibfnamefont {M.}~\bibnamefont {Liebend{\"o}rfer}}, \bibinfo {author}
  {\bibfnamefont {H.~T.}\ \bibnamefont {Janka}}, \ and\ \bibinfo {author}
  {\bibfnamefont {M.}~\bibnamefont {Rampp}},\ }\href {\doibase
  10.1103/PhysRevLett.90.241102} {\bibfield  {journal} {\bibinfo  {journal}
  {Phys. Rev. Lett.}\ }\textbf {\bibinfo {volume} {90}},\ \bibinfo {pages}
  {241102} (\bibinfo {year} {2003})}\BibitemShut {NoStop}%
\bibitem [{\citenamefont {Typel}\ \emph {et~al.}(2015)\citenamefont {Typel},
  \citenamefont {Oertel},\ and\ \citenamefont {Kl{\"a}hn}}]{Typel2013}%
  \BibitemOpen
  \bibfield  {author} {\bibinfo {author} {\bibfnamefont {S.}~\bibnamefont
  {Typel}}, \bibinfo {author} {\bibfnamefont {M.}~\bibnamefont {Oertel}}, \
  and\ \bibinfo {author} {\bibfnamefont {T.}~\bibnamefont {Kl{\"a}hn}},\ }\href
  {\doibase 10.1134/S1063779615040061} {\bibfield  {journal} {\bibinfo
  {journal} {Phys. Part. Nucl.}\ }\textbf {\bibinfo {volume} {46}},\ \bibinfo
  {pages} {633} (\bibinfo {year} {2015})}\BibitemShut {NoStop}%
\bibitem [{\citenamefont {{Dimmelmeier}}\ \emph {et~al.}(2012)\citenamefont
  {{Dimmelmeier}}, \citenamefont {{Novak}},\ and\ \citenamefont
  {{Cerd{\'a}-Dur{\'a}n}}}]{coconut}%
  \BibitemOpen
  \bibfield  {author} {\bibinfo {author} {\bibfnamefont {H.}~\bibnamefont
  {{Dimmelmeier}}}, \bibinfo {author} {\bibfnamefont {J.}~\bibnamefont
  {{Novak}}}, \ and\ \bibinfo {author} {\bibfnamefont {P.}~\bibnamefont
  {{Cerd{\'a}-Dur{\'a}n}}},\ }\href@noop {} {\enquote {\bibinfo {title}
  {{CoCoNuT: General relativistic hydrodynamics code with dynamical space- time
  evolution}},}\ }\bibinfo {howpublished} {Astrophysics Source Code Library}
  (\bibinfo {year} {2012}),\ \Eprint {http://arxiv.org/abs/1202.012}
  {ascl:1202.012} \BibitemShut {NoStop}%
\bibitem [{\citenamefont {Dimmelmeier}\ \emph {et~al.}(2005)\citenamefont
  {Dimmelmeier}, \citenamefont {Novak}, \citenamefont {Font}, \citenamefont
  {Ib{\'a}{\~n}ez},\ and\ \citenamefont {M{\"u}ller}}]{dimmelmeier-05}%
  \BibitemOpen
  \bibfield  {author} {\bibinfo {author} {\bibfnamefont {H.}~\bibnamefont
  {Dimmelmeier}}, \bibinfo {author} {\bibfnamefont {J.}~\bibnamefont {Novak}},
  \bibinfo {author} {\bibfnamefont {J.~A.}\ \bibnamefont {Font}}, \bibinfo
  {author} {\bibfnamefont {J.~M.}\ \bibnamefont {Ib{\'a}{\~n}ez}}, \ and\
  \bibinfo {author} {\bibfnamefont {E.}~\bibnamefont {M{\"u}ller}},\ }\href
  {\doibase 10.1103/PhysRevD.71.064023} {\bibfield  {journal} {\bibinfo
  {journal} {Phys. Rev. D}\ }\textbf {\bibinfo {volume} {71}},\ \bibinfo
  {pages} {1} (\bibinfo {year} {2005})}\BibitemShut {NoStop}%
\bibitem [{\citenamefont {{M{\"u}ller}}\ and\ \citenamefont
  {{Janka}}(2015)}]{mueller-15}%
  \BibitemOpen
  \bibfield  {author} {\bibinfo {author} {\bibfnamefont {B.}~\bibnamefont
  {{M{\"u}ller}}}\ and\ \bibinfo {author} {\bibfnamefont {H.~T.}\ \bibnamefont
  {{Janka}}},\ }\href {\doibase 10.1093/mnras/stv101} {\bibfield  {journal}
  {\bibinfo  {journal} {\mnras}\ }\textbf {\bibinfo {volume} {448}},\ \bibinfo
  {pages} {2141} (\bibinfo {year} {2015})},\ \Eprint
  {http://arxiv.org/abs/1409.4783} {arXiv:1409.4783 [astro-ph.SR]} \BibitemShut
  {NoStop}%
\bibitem [{\citenamefont {Grams}\ \emph {et~al.}(2018)\citenamefont {Grams},
  \citenamefont {Giraud}, \citenamefont {Fantina},\ and\ \citenamefont
  {Gulminelli}}]{grams2018}%
  \BibitemOpen
  \bibfield  {author} {\bibinfo {author} {\bibfnamefont {G.}~\bibnamefont
  {Grams}}, \bibinfo {author} {\bibfnamefont {S.}~\bibnamefont {Giraud}},
  \bibinfo {author} {\bibfnamefont {A.~F.}\ \bibnamefont {Fantina}}, \ and\
  \bibinfo {author} {\bibfnamefont {F.}~\bibnamefont {Gulminelli}},\ }\href
  {\doibase 10.1103/PhysRevC.97.035807} {\bibfield  {journal} {\bibinfo
  {journal} {Phys. Rev.}\ }\textbf {\bibinfo {volume} {C97}},\ \bibinfo {pages}
  {035807} (\bibinfo {year} {2018})}\BibitemShut {NoStop}%
\bibitem [{\citenamefont {Romero}\ \emph {et~al.}(1996)\citenamefont {Romero},
  \citenamefont {Ibanez}, \citenamefont {Marti},\ and\ \citenamefont
  {Miralles}}]{Romero1996}%
  \BibitemOpen
  \bibfield  {author} {\bibinfo {author} {\bibfnamefont {J.~V.}\ \bibnamefont
  {Romero}}, \bibinfo {author} {\bibfnamefont {J.~M.}\ \bibnamefont {Ibanez}},
  \bibinfo {author} {\bibfnamefont {J.~M.}\ \bibnamefont {Marti}}, \ and\
  \bibinfo {author} {\bibfnamefont {J.~A.}\ \bibnamefont {Miralles}},\ }\href
  {\doibase 10.1086/177198} {\bibfield  {journal} {\bibinfo  {journal}
  {Astrophys. J.}\ }\textbf {\bibinfo {volume} {462}},\ \bibinfo {pages} {839}
  (\bibinfo {year} {1996})},\ \Eprint {http://arxiv.org/abs/astro-ph/9509121}
  {arXiv:astro-ph/9509121 [astro-ph]} \BibitemShut {NoStop}%
\bibitem [{\citenamefont {{Romero}}\ \emph {et~al.}(1997)\citenamefont
  {{Romero}}, \citenamefont {{Ib\'a\~{n}ez}}, \citenamefont {{Miralles}},\ and\
  \citenamefont {{Pons}}}]{romero1996b}%
  \BibitemOpen
  \bibfield  {author} {\bibinfo {author} {\bibfnamefont {J.~V.}\ \bibnamefont
  {{Romero}}}, \bibinfo {author} {\bibfnamefont {J.~M.}\ \bibnamefont
  {{Ib\'a\~{n}ez}}}, \bibinfo {author} {\bibfnamefont {J.~A.}\ \bibnamefont
  {{Miralles}}}, \ and\ \bibinfo {author} {\bibfnamefont {J.~A.}\ \bibnamefont
  {{Pons}}},\ }in\ \href@noop {} {\emph {\bibinfo {booktitle} {Some Topics on
  general relativity and gravitational radiation}}},\ \bibinfo {editor} {edited
  by\ \bibinfo {editor} {\bibfnamefont {J.~A.}\ \bibnamefont {{Miralles}}},
  \bibinfo {editor} {\bibfnamefont {J.~A.}\ \bibnamefont {{Morales}}}, \ and\
  \bibinfo {editor} {\bibfnamefont {D.}~\bibnamefont {{S\'aez}}}}\ (\bibinfo
  {year} {1997})\ p.\ \bibinfo {pages} {289}\BibitemShut {NoStop}%
\bibitem [{\citenamefont {{Fantina}}(2010)}]{fantinaphd}%
  \BibitemOpen
  \bibfield  {author} {\bibinfo {author} {\bibfnamefont {A.~F.}\ \bibnamefont
  {{Fantina}}},\ }\emph {\bibinfo {title} {{Supernovae theory: study of
  electro-weak processes during gravitational collapse of massive stars}}},\
  \href@noop {} {Ph.D. thesis},\ \bibinfo  {school} {Universi\'e Paris XI,
  Orsay, France, and Universit\`a degli Studi di Milano, Milano, Italy}
  (\bibinfo {year} {2010})\BibitemShut {NoStop}%
\bibitem [{\citenamefont {{Woosley}}\ \emph {et~al.}(2002)\citenamefont
  {{Woosley}}, \citenamefont {{Heger}},\ and\ \citenamefont
  {{Weaver}}}]{woosley-02}%
  \BibitemOpen
  \bibfield  {author} {\bibinfo {author} {\bibfnamefont {S.~E.}\ \bibnamefont
  {{Woosley}}}, \bibinfo {author} {\bibfnamefont {A.}~\bibnamefont {{Heger}}},
  \ and\ \bibinfo {author} {\bibfnamefont {T.~A.}\ \bibnamefont {{Weaver}}},\
  }\href {\doibase 10.1103/RevModPhys.74.1015} {\bibfield  {journal} {\bibinfo
  {journal} {Rev. Mod. Phys.}\ }\textbf {\bibinfo {volume} {74}},\ \bibinfo
  {pages} {1015} (\bibinfo {year} {2002})}\BibitemShut {NoStop}%
\bibitem [{\citenamefont {Lattimer}\ and\ \citenamefont
  {Swesty}(1991)}]{Lattimer_NPA_1991}%
  \BibitemOpen
  \bibfield  {author} {\bibinfo {author} {\bibfnamefont {J.~M.}\ \bibnamefont
  {Lattimer}}\ and\ \bibinfo {author} {\bibfnamefont {F.~D.}\ \bibnamefont
  {Swesty}},\ }\href {\doibase https://doi.org/10.1016/0375-9474(91)90452-C}
  {\bibfield  {journal} {\bibinfo  {journal} {Nuclear Physics A}\ }\textbf
  {\bibinfo {volume} {535}},\ \bibinfo {pages} {331 } (\bibinfo {year}
  {1991})}\BibitemShut {NoStop}%
\bibitem [{\citenamefont {Shen}\ \emph {et~al.}(1998)\citenamefont {Shen},
  \citenamefont {Toki}, \citenamefont {Oyamatsu},\ and\ \citenamefont
  {Sumiyoshi}}]{Shen_NPA_1998}%
  \BibitemOpen
  \bibfield  {author} {\bibinfo {author} {\bibfnamefont {H.}~\bibnamefont
  {Shen}}, \bibinfo {author} {\bibfnamefont {H.}~\bibnamefont {Toki}}, \bibinfo
  {author} {\bibfnamefont {K.}~\bibnamefont {Oyamatsu}}, \ and\ \bibinfo
  {author} {\bibfnamefont {K.}~\bibnamefont {Sumiyoshi}},\ }\href {\doibase
  https://doi.org/10.1016/S0375-9474(98)00236-X} {\bibfield  {journal}
  {\bibinfo  {journal} {Nuclear Physics A}\ }\textbf {\bibinfo {volume}
  {637}},\ \bibinfo {pages} {435 } (\bibinfo {year} {1998})}\BibitemShut
  {NoStop}%
\bibitem [{\citenamefont {Lattimer}\ \emph {et~al.}(1985)\citenamefont
  {Lattimer}, \citenamefont {Pethick}, \citenamefont {Ravenhall},\ and\
  \citenamefont {Lamb}}]{Lattimer_NPA_1985}%
  \BibitemOpen
  \bibfield  {author} {\bibinfo {author} {\bibfnamefont {J.}~\bibnamefont
  {Lattimer}}, \bibinfo {author} {\bibfnamefont {C.}~\bibnamefont {Pethick}},
  \bibinfo {author} {\bibfnamefont {D.}~\bibnamefont {Ravenhall}}, \ and\
  \bibinfo {author} {\bibfnamefont {D.}~\bibnamefont {Lamb}},\ }\href {\doibase
  https://doi.org/10.1016/0375-9474(85)90006-5} {\bibfield  {journal} {\bibinfo
   {journal} {Nuclear Physics A}\ }\textbf {\bibinfo {volume} {432}},\ \bibinfo
  {pages} {646 } (\bibinfo {year} {1985})}\BibitemShut {NoStop}%
\bibitem [{\citenamefont {Hempel}\ \emph {et~al.}(2012)\citenamefont {Hempel},
  \citenamefont {Fischer}, \citenamefont {Schaffner-Bielich},\ and\
  \citenamefont {Liebendörfer}}]{Hempel_ApJ_2012}%
  \BibitemOpen
  \bibfield  {author} {\bibinfo {author} {\bibfnamefont {M.}~\bibnamefont
  {Hempel}}, \bibinfo {author} {\bibfnamefont {T.}~\bibnamefont {Fischer}},
  \bibinfo {author} {\bibfnamefont {J.}~\bibnamefont {Schaffner-Bielich}}, \
  and\ \bibinfo {author} {\bibfnamefont {M.}~\bibnamefont {Liebendörfer}},\
  }\href {http://stacks.iop.org/0004-637X/748/i=1/a=70} {\bibfield  {journal}
  {\bibinfo  {journal} {The Astrophysical Journal}\ }\textbf {\bibinfo {volume}
  {748}},\ \bibinfo {pages} {70} (\bibinfo {year} {2012})}\BibitemShut
  {NoStop}%
\bibitem [{\citenamefont {Hempel}\ and\ \citenamefont
  {Schaffner-Bielich}(2010)}]{Hempel_NPA_2010}%
  \BibitemOpen
  \bibfield  {author} {\bibinfo {author} {\bibfnamefont {M.}~\bibnamefont
  {Hempel}}\ and\ \bibinfo {author} {\bibfnamefont {J.}~\bibnamefont
  {Schaffner-Bielich}},\ }\href {\doibase
  https://doi.org/10.1016/j.nuclphysa.2010.02.010} {\bibfield  {journal}
  {\bibinfo  {journal} {Nuclear Physics A}\ }\textbf {\bibinfo {volume}
  {837}},\ \bibinfo {pages} {210 } (\bibinfo {year} {2010})}\BibitemShut
  {NoStop}%
\bibitem [{\citenamefont {Gulminelli}\ and\ \citenamefont
  {Raduta}(2015)}]{Gulminelli_PRC_2015}%
  \BibitemOpen
  \bibfield  {author} {\bibinfo {author} {\bibfnamefont {F.}~\bibnamefont
  {Gulminelli}}\ and\ \bibinfo {author} {\bibfnamefont {A.~R.}\ \bibnamefont
  {Raduta}},\ }\href {\doibase 10.1103/PhysRevC.92.055803} {\bibfield
  {journal} {\bibinfo  {journal} {Phys. Rev. C}\ }\textbf {\bibinfo {volume}
  {92}},\ \bibinfo {pages} {055803} (\bibinfo {year} {2015})}\BibitemShut
  {NoStop}%
\bibitem [{\citenamefont {Raduta}\ and\ \citenamefont
  {Gulminelli}(2019)}]{Raduta_2019}%
  \BibitemOpen
  \bibfield  {author} {\bibinfo {author} {\bibfnamefont {A.~R.}\ \bibnamefont
  {Raduta}}\ and\ \bibinfo {author} {\bibfnamefont {F.}~\bibnamefont
  {Gulminelli}},\ }\href {\doibase 10.1016/j.nuclphysa.2018.11.003} {\bibfield
  {journal} {\bibinfo  {journal} {Nucl. Phys.}\ }\textbf {\bibinfo {volume}
  {A983}},\ \bibinfo {pages} {252} (\bibinfo {year} {2019})}\BibitemShut
  {NoStop}%
\bibitem [{\citenamefont {{Furusawa}}\ \emph {et~al.}(2017)\citenamefont
  {{Furusawa}}, \citenamefont {{Togashi}}, \citenamefont {{Nagakura}},
  \citenamefont {{Sumiyoshi}}, \citenamefont {{Yamada}}, \citenamefont
  {{Suzuki}},\ and\ \citenamefont {{Takano}}}]{TN_2017}%
  \BibitemOpen
  \bibfield  {author} {\bibinfo {author} {\bibfnamefont {S.}~\bibnamefont
  {{Furusawa}}}, \bibinfo {author} {\bibfnamefont {H.}~\bibnamefont
  {{Togashi}}}, \bibinfo {author} {\bibfnamefont {H.}~\bibnamefont
  {{Nagakura}}}, \bibinfo {author} {\bibfnamefont {K.}~\bibnamefont
  {{Sumiyoshi}}}, \bibinfo {author} {\bibfnamefont {S.}~\bibnamefont
  {{Yamada}}}, \bibinfo {author} {\bibfnamefont {H.}~\bibnamefont {{Suzuki}}},
  \ and\ \bibinfo {author} {\bibfnamefont {M.}~\bibnamefont {{Takano}}},\
  }\href {\doibase 10.1088/1361-6471/aa7f35} {\bibfield  {journal} {\bibinfo
  {journal} {Journal of Physics G Nuclear Physics}\ }\textbf {\bibinfo {volume}
  {44}},\ \bibinfo {pages} {094001} (\bibinfo {year} {2017})}\BibitemShut
  {NoStop}%
\bibitem [{\citenamefont {Furusawa}\ \emph {et~al.}(2017)\citenamefont
  {Furusawa}, \citenamefont {Sumiyoshi}, \citenamefont {Yamada},\ and\
  \citenamefont {Suzuki}}]{FYSS_2017}%
  \BibitemOpen
  \bibfield  {author} {\bibinfo {author} {\bibfnamefont {S.}~\bibnamefont
  {Furusawa}}, \bibinfo {author} {\bibfnamefont {K.}~\bibnamefont {Sumiyoshi}},
  \bibinfo {author} {\bibfnamefont {S.}~\bibnamefont {Yamada}}, \ and\ \bibinfo
  {author} {\bibfnamefont {H.}~\bibnamefont {Suzuki}},\ }\href {\doibase
  https://doi.org/10.1016/j.nuclphysa.2016.09.002} {\bibfield  {journal}
  {\bibinfo  {journal} {Nuclear Physics A}\ }\textbf {\bibinfo {volume}
  {957}},\ \bibinfo {pages} {188 } (\bibinfo {year} {2017})}\BibitemShut
  {NoStop}%
\bibitem [{\citenamefont {Typel}(2018)}]{Typel_2018}%
  \BibitemOpen
  \bibfield  {author} {\bibinfo {author} {\bibfnamefont {S.}~\bibnamefont
  {Typel}},\ }\href {\doibase 10.1088/1361-6471/aadea5} {\bibfield  {journal}
  {\bibinfo  {journal} {J. Phys.}\ }\textbf {\bibinfo {volume} {G45}},\
  \bibinfo {pages} {114001} (\bibinfo {year} {2018})}\BibitemShut {NoStop}%
\bibitem [{\citenamefont {Typel}\ \emph {et~al.}(2010)\citenamefont {Typel},
  \citenamefont {R\"opke}, \citenamefont {Kl\"ahn}, \citenamefont {Blaschke},\
  and\ \citenamefont {Wolter}}]{DD2}%
  \BibitemOpen
  \bibfield  {author} {\bibinfo {author} {\bibfnamefont {S.}~\bibnamefont
  {Typel}}, \bibinfo {author} {\bibfnamefont {G.}~\bibnamefont {R\"opke}},
  \bibinfo {author} {\bibfnamefont {T.}~\bibnamefont {Kl\"ahn}}, \bibinfo
  {author} {\bibfnamefont {D.}~\bibnamefont {Blaschke}}, \ and\ \bibinfo
  {author} {\bibfnamefont {H.~H.}\ \bibnamefont {Wolter}},\ }\href {\doibase
  10.1103/PhysRevC.81.015803} {\bibfield  {journal} {\bibinfo  {journal} {Phys.
  Rev. C}\ }\textbf {\bibinfo {volume} {81}},\ \bibinfo {pages} {015803}
  (\bibinfo {year} {2010})}\BibitemShut {NoStop}%
\bibitem [{\citenamefont {Moller}\ \emph {et~al.}(1995)\citenamefont {Moller},
  \citenamefont {Nix}, \citenamefont {Myers},\ and\ \citenamefont
  {Swiatecki}}]{FRDM}%
  \BibitemOpen
  \bibfield  {author} {\bibinfo {author} {\bibfnamefont {P.}~\bibnamefont
  {Moller}}, \bibinfo {author} {\bibfnamefont {J.}~\bibnamefont {Nix}},
  \bibinfo {author} {\bibfnamefont {W.}~\bibnamefont {Myers}}, \ and\ \bibinfo
  {author} {\bibfnamefont {W.}~\bibnamefont {Swiatecki}},\ }\href {\doibase
  https://doi.org/10.1006/adnd.1995.1002} {\bibfield  {journal} {\bibinfo
  {journal} {Atomic Data and Nuclear Data Tables}\ }\textbf {\bibinfo {volume}
  {59}},\ \bibinfo {pages} {185 } (\bibinfo {year} {1995})}\BibitemShut
  {NoStop}%
\bibitem [{\citenamefont {Chabanat}\ \emph {et~al.}(1998)\citenamefont
  {Chabanat}, \citenamefont {Bonche}, \citenamefont {Haensel}, \citenamefont
  {Meyer},\ and\ \citenamefont {Schaeffer}}]{SLy4}%
  \BibitemOpen
  \bibfield  {author} {\bibinfo {author} {\bibfnamefont {E.}~\bibnamefont
  {Chabanat}}, \bibinfo {author} {\bibfnamefont {P.}~\bibnamefont {Bonche}},
  \bibinfo {author} {\bibfnamefont {P.}~\bibnamefont {Haensel}}, \bibinfo
  {author} {\bibfnamefont {J.}~\bibnamefont {Meyer}}, \ and\ \bibinfo {author}
  {\bibfnamefont {R.}~\bibnamefont {Schaeffer}},\ }\href {\doibase
  https://doi.org/10.1016/S0375-9474(98)00180-8} {\bibfield  {journal}
  {\bibinfo  {journal} {Nuclear Physics A}\ }\textbf {\bibinfo {volume}
  {635}},\ \bibinfo {pages} {231 } (\bibinfo {year} {1998})}\BibitemShut
  {NoStop}%
\bibitem [{\citenamefont {Audi}\ \emph {et~al.}(2012)\citenamefont {Audi},
  \citenamefont {Kondev}, \citenamefont {Wang}, \citenamefont {Pfeiffer},
  \citenamefont {Sun}, \citenamefont {Blachot},\ and\ \citenamefont
  {MacCormick}}]{AME2012a}%
  \BibitemOpen
  \bibfield  {author} {\bibinfo {author} {\bibfnamefont {G.}~\bibnamefont
  {Audi}}, \bibinfo {author} {\bibfnamefont {F.}~\bibnamefont {Kondev}},
  \bibinfo {author} {\bibfnamefont {M.}~\bibnamefont {Wang}}, \bibinfo {author}
  {\bibfnamefont {B.}~\bibnamefont {Pfeiffer}}, \bibinfo {author}
  {\bibfnamefont {X.}~\bibnamefont {Sun}}, \bibinfo {author} {\bibfnamefont
  {J.}~\bibnamefont {Blachot}}, \ and\ \bibinfo {author} {\bibfnamefont
  {M.}~\bibnamefont {MacCormick}},\ }\href
  {http://stacks.iop.org/1674-1137/36/i=12/a=001} {\bibfield  {journal}
  {\bibinfo  {journal} {Chinese Physics C}\ }\textbf {\bibinfo {volume} {36}},\
  \bibinfo {pages} {1157} (\bibinfo {year} {2012})}\BibitemShut {NoStop}%
\bibitem [{\citenamefont {Wang}\ \emph {et~al.}(2012)\citenamefont {Wang},
  \citenamefont {Audi}, \citenamefont {Wapstra}, \citenamefont {Kondev},
  \citenamefont {MacCormick}, \citenamefont {Xu},\ and\ \citenamefont
  {Pfeiffer}}]{AME2012b}%
  \BibitemOpen
  \bibfield  {author} {\bibinfo {author} {\bibfnamefont {M.}~\bibnamefont
  {Wang}}, \bibinfo {author} {\bibfnamefont {G.}~\bibnamefont {Audi}}, \bibinfo
  {author} {\bibfnamefont {A.}~\bibnamefont {Wapstra}}, \bibinfo {author}
  {\bibfnamefont {F.}~\bibnamefont {Kondev}}, \bibinfo {author} {\bibfnamefont
  {M.}~\bibnamefont {MacCormick}}, \bibinfo {author} {\bibfnamefont
  {X.}~\bibnamefont {Xu}}, \ and\ \bibinfo {author} {\bibfnamefont
  {B.}~\bibnamefont {Pfeiffer}},\ }\href
  {http://stacks.iop.org/1674-1137/36/i=12/a=003} {\bibfield  {journal}
  {\bibinfo  {journal} {Chinese Physics C}\ }\textbf {\bibinfo {volume} {36}},\
  \bibinfo {pages} {1603} (\bibinfo {year} {2012})}\BibitemShut {NoStop}%
\bibitem [{\citenamefont {Duflo}\ and\ \citenamefont {Zuker}(1995)}]{DZ10}%
  \BibitemOpen
  \bibfield  {author} {\bibinfo {author} {\bibfnamefont {J.}~\bibnamefont
  {Duflo}}\ and\ \bibinfo {author} {\bibfnamefont {A.}~\bibnamefont {Zuker}},\
  }\href {\doibase 10.1103/PhysRevC.52.R23} {\bibfield  {journal} {\bibinfo
  {journal} {Phys. Rev. C}\ }\textbf {\bibinfo {volume} {52}},\ \bibinfo
  {pages} {R23} (\bibinfo {year} {1995})}\BibitemShut {NoStop}%
\bibitem [{\citenamefont {Danielewicz}\ and\ \citenamefont
  {Lee}(2009)}]{Danielewicz_NPA_2009}%
  \BibitemOpen
  \bibfield  {author} {\bibinfo {author} {\bibfnamefont {P.}~\bibnamefont
  {Danielewicz}}\ and\ \bibinfo {author} {\bibfnamefont {J.}~\bibnamefont
  {Lee}},\ }\href {\doibase https://doi.org/10.1016/j.nuclphysa.2008.11.007}
  {\bibfield  {journal} {\bibinfo  {journal} {Nuclear Physics A}\ }\textbf
  {\bibinfo {volume} {818}},\ \bibinfo {pages} {36 } (\bibinfo {year}
  {2009})}\BibitemShut {NoStop}%
\bibitem [{\citenamefont {Xu}\ \emph {et~al.}(2013)\citenamefont {Xu},
  \citenamefont {Goriely}, \citenamefont {Jorissen}, \citenamefont {Chen},\
  and\ \citenamefont {Arnould}}]{xu2013}%
  \BibitemOpen
  \bibfield  {author} {\bibinfo {author} {\bibfnamefont {Y.}~\bibnamefont
  {Xu}}, \bibinfo {author} {\bibfnamefont {S.}~\bibnamefont {Goriely}},
  \bibinfo {author} {\bibfnamefont {A.}~\bibnamefont {Jorissen}}, \bibinfo
  {author} {\bibfnamefont {G.}~\bibnamefont {Chen}}, \ and\ \bibinfo {author}
  {\bibfnamefont {M.}~\bibnamefont {Arnould}},\ }\href {\doibase
  10.1051/0004-6361/201220537} {\bibfield  {journal} {\bibinfo  {journal}
  {Astron. Astrophys.}\ }\textbf {\bibinfo {volume} {549}},\ \bibinfo {pages}
  {A106} (\bibinfo {year} {2013})},\ \Eprint {http://arxiv.org/abs/1212.0628}
  {arXiv:1212.0628 [nucl-th]} \BibitemShut {NoStop}%
\bibitem [{\citenamefont {{Goriely}}\ \emph {et~al.}(2013)\citenamefont
  {{Goriely}}, \citenamefont {{Chamel}},\ and\ \citenamefont
  {{Pearson}}}]{gcp2013}%
  \BibitemOpen
  \bibfield  {author} {\bibinfo {author} {\bibfnamefont {S.}~\bibnamefont
  {{Goriely}}}, \bibinfo {author} {\bibfnamefont {N.}~\bibnamefont {{Chamel}}},
  \ and\ \bibinfo {author} {\bibfnamefont {J.~M.}\ \bibnamefont {{Pearson}}},\
  }\href {\doibase 10.1103/PhysRevC.88.024308} {\bibfield  {journal} {\bibinfo
  {journal} {\prc}\ }\textbf {\bibinfo {volume} {88}},\ \bibinfo {eid} {024308}
  (\bibinfo {year} {2013})}\BibitemShut {NoStop}%
\bibitem [{\citenamefont {{Fuller}}\ \emph {et~al.}(1980)\citenamefont
  {{Fuller}}, \citenamefont {{Fowler}},\ and\ \citenamefont
  {{Newman}}}]{FFN_1980}%
  \BibitemOpen
  \bibfield  {author} {\bibinfo {author} {\bibfnamefont {G.~M.}\ \bibnamefont
  {{Fuller}}}, \bibinfo {author} {\bibfnamefont {W.~A.}\ \bibnamefont
  {{Fowler}}}, \ and\ \bibinfo {author} {\bibfnamefont {M.~J.}\ \bibnamefont
  {{Newman}}},\ }\href {\doibase 10.1086/190657} {\bibfield  {journal}
  {\bibinfo  {journal} {Astrophys. J. Suppl.}\ }\textbf {\bibinfo {volume}
  {42}},\ \bibinfo {pages} {447} (\bibinfo {year} {1980})}\BibitemShut
  {NoStop}%
\bibitem [{\citenamefont {{Fuller}}\ \emph
  {et~al.}(1982{\natexlab{b}})\citenamefont {{Fuller}}, \citenamefont
  {{Fowler}},\ and\ \citenamefont {{Newman}}}]{FFN_1982a}%
  \BibitemOpen
  \bibfield  {author} {\bibinfo {author} {\bibfnamefont {G.~M.}\ \bibnamefont
  {{Fuller}}}, \bibinfo {author} {\bibfnamefont {W.~A.}\ \bibnamefont
  {{Fowler}}}, \ and\ \bibinfo {author} {\bibfnamefont {M.~J.}\ \bibnamefont
  {{Newman}}},\ }\href {\doibase 10.1086/159597} {\bibfield  {journal}
  {\bibinfo  {journal} {Astrophys. J.}\ }\textbf {\bibinfo {volume} {252}},\
  \bibinfo {pages} {715} (\bibinfo {year} {1982}{\natexlab{b}})}\BibitemShut
  {NoStop}%
\bibitem [{\citenamefont {{Fuller}}\ \emph {et~al.}(1985)\citenamefont
  {{Fuller}}, \citenamefont {{Fowler}},\ and\ \citenamefont
  {{Newman}}}]{FFN_1985}%
  \BibitemOpen
  \bibfield  {author} {\bibinfo {author} {\bibfnamefont {G.~M.}\ \bibnamefont
  {{Fuller}}}, \bibinfo {author} {\bibfnamefont {W.~A.}\ \bibnamefont
  {{Fowler}}}, \ and\ \bibinfo {author} {\bibfnamefont {M.~J.}\ \bibnamefont
  {{Newman}}},\ }\href {\doibase 10.1086/163208} {\bibfield  {journal}
  {\bibinfo  {journal} {Astrophys. J.}\ }\textbf {\bibinfo {volume} {293}},\
  \bibinfo {pages} {1} (\bibinfo {year} {1985})}\BibitemShut {NoStop}%
\bibitem [{\citenamefont {Langanke}\ and\ \citenamefont
  {Martıinez-Pinedo}(2000)}]{LMP_NPA_2000}%
  \BibitemOpen
  \bibfield  {author} {\bibinfo {author} {\bibfnamefont {K.}~\bibnamefont
  {Langanke}}\ and\ \bibinfo {author} {\bibfnamefont {G.}~\bibnamefont
  {Martıinez-Pinedo}},\ }\href {\doibase
  https://doi.org/10.1016/S0375-9474(00)00131-7} {\bibfield  {journal}
  {\bibinfo  {journal} {Nuclear Physics A}\ }\textbf {\bibinfo {volume}
  {673}},\ \bibinfo {pages} {481 } (\bibinfo {year} {2000})}\BibitemShut
  {NoStop}%
\bibitem [{\citenamefont {Paar}\ \emph {et~al.}(2009)\citenamefont {Paar},
  \citenamefont {Col\`o}, \citenamefont {Khan},\ and\ \citenamefont
  {Vretenar}}]{Paar_PRC_2009}%
  \BibitemOpen
  \bibfield  {author} {\bibinfo {author} {\bibfnamefont {N.}~\bibnamefont
  {Paar}}, \bibinfo {author} {\bibfnamefont {G.}~\bibnamefont {Col\`o}},
  \bibinfo {author} {\bibfnamefont {E.}~\bibnamefont {Khan}}, \ and\ \bibinfo
  {author} {\bibfnamefont {D.}~\bibnamefont {Vretenar}},\ }\href {\doibase
  10.1103/PhysRevC.80.055801} {\bibfield  {journal} {\bibinfo  {journal} {Phys.
  Rev. C}\ }\textbf {\bibinfo {volume} {80}},\ \bibinfo {pages} {055801}
  (\bibinfo {year} {2009})}\BibitemShut {NoStop}%
\bibitem [{\citenamefont {Niu}\ \emph {et~al.}(2011)\citenamefont {Niu},
  \citenamefont {Paar}, \citenamefont {Vretenar},\ and\ \citenamefont
  {Meng}}]{Niu_PRC_2011}%
  \BibitemOpen
  \bibfield  {author} {\bibinfo {author} {\bibfnamefont {Y.~F.}\ \bibnamefont
  {Niu}}, \bibinfo {author} {\bibfnamefont {N.}~\bibnamefont {Paar}}, \bibinfo
  {author} {\bibfnamefont {D.}~\bibnamefont {Vretenar}}, \ and\ \bibinfo
  {author} {\bibfnamefont {J.}~\bibnamefont {Meng}},\ }\href {\doibase
  10.1103/PhysRevC.83.045807} {\bibfield  {journal} {\bibinfo  {journal} {Phys.
  Rev. C}\ }\textbf {\bibinfo {volume} {83}},\ \bibinfo {pages} {045807}
  (\bibinfo {year} {2011})}\BibitemShut {NoStop}%
\bibitem [{\citenamefont {Nabi}\ and\ \citenamefont
  {Klapdor-Kleingrothaus}(1999)}]{Nabi_1999}%
  \BibitemOpen
  \bibfield  {author} {\bibinfo {author} {\bibfnamefont {J.-U.}\ \bibnamefont
  {Nabi}}\ and\ \bibinfo {author} {\bibfnamefont {H.}~\bibnamefont
  {Klapdor-Kleingrothaus}},\ }\href {\doibase
  https://doi.org/10.1006/adnd.1998.0801} {\bibfield  {journal} {\bibinfo
  {journal} {Atomic Data and Nuclear Data Tables}\ }\textbf {\bibinfo {volume}
  {71}},\ \bibinfo {pages} {149 } (\bibinfo {year} {1999})}\BibitemShut
  {NoStop}%
\bibitem [{\citenamefont {Nabi}\ and\ \citenamefont
  {Klapdor-Kleingrothaus}(2004)}]{Nabi_2004}%
  \BibitemOpen
  \bibfield  {author} {\bibinfo {author} {\bibfnamefont {J.-U.}\ \bibnamefont
  {Nabi}}\ and\ \bibinfo {author} {\bibfnamefont {H.~V.}\ \bibnamefont
  {Klapdor-Kleingrothaus}},\ }\href {\doibase
  https://doi.org/10.1016/j.adt.2004.09.002} {\bibfield  {journal} {\bibinfo
  {journal} {Atomic Data and Nuclear Data Tables}\ }\textbf {\bibinfo {volume}
  {88}},\ \bibinfo {pages} {237 } (\bibinfo {year} {2004})}\BibitemShut
  {NoStop}%
\bibitem [{\citenamefont {Caurier}\ \emph {et~al.}(1999)\citenamefont
  {Caurier}, \citenamefont {Langanke}, \citenamefont {Martínez-Pinedo},\ and\
  \citenamefont {Nowacki}}]{Caurier_1999}%
  \BibitemOpen
  \bibfield  {author} {\bibinfo {author} {\bibfnamefont {E.}~\bibnamefont
  {Caurier}}, \bibinfo {author} {\bibfnamefont {K.}~\bibnamefont {Langanke}},
  \bibinfo {author} {\bibfnamefont {G.}~\bibnamefont {Martínez-Pinedo}}, \
  and\ \bibinfo {author} {\bibfnamefont {F.}~\bibnamefont {Nowacki}},\ }\href
  {\doibase https://doi.org/10.1016/S0375-9474(99)00240-7} {\bibfield
  {journal} {\bibinfo  {journal} {Nuclear Physics A}\ }\textbf {\bibinfo
  {volume} {653}},\ \bibinfo {pages} {439 } (\bibinfo {year}
  {1999})}\BibitemShut {NoStop}%
\bibitem [{\citenamefont {Dean}\ \emph {et~al.}(1998)\citenamefont {Dean},
  \citenamefont {Langanke}, \citenamefont {Chatterjee}, \citenamefont {Radha},\
  and\ \citenamefont {Strayer}}]{Dean_PRC_1998}%
  \BibitemOpen
  \bibfield  {author} {\bibinfo {author} {\bibfnamefont {D.~J.}\ \bibnamefont
  {Dean}}, \bibinfo {author} {\bibfnamefont {K.}~\bibnamefont {Langanke}},
  \bibinfo {author} {\bibfnamefont {L.}~\bibnamefont {Chatterjee}}, \bibinfo
  {author} {\bibfnamefont {P.~B.}\ \bibnamefont {Radha}}, \ and\ \bibinfo
  {author} {\bibfnamefont {M.~R.}\ \bibnamefont {Strayer}},\ }\href {\doibase
  10.1103/PhysRevC.58.536} {\bibfield  {journal} {\bibinfo  {journal} {Phys.
  Rev. C}\ }\textbf {\bibinfo {volume} {58}},\ \bibinfo {pages} {536} (\bibinfo
  {year} {1998})}\BibitemShut {NoStop}%
\bibitem [{\citenamefont {{Bruenn}}(1985)}]{Bruenn_1985}%
  \BibitemOpen
  \bibfield  {author} {\bibinfo {author} {\bibfnamefont {S.~W.}\ \bibnamefont
  {{Bruenn}}},\ }\href {\doibase 10.1086/191056} {\bibfield  {journal}
  {\bibinfo  {journal} {Astrophys. J. Suppl.}\ }\textbf {\bibinfo {volume}
  {58}},\ \bibinfo {pages} {771} (\bibinfo {year} {1985})}\BibitemShut
  {NoStop}%
\bibitem [{\citenamefont {Juodagalvis}\ \emph {et~al.}(2007)\citenamefont
  {Juodagalvis}, \citenamefont {Sampaio}, \citenamefont {Langanke},\ and\
  \citenamefont {Hix}}]{Juodagalvis_JPG_2007}%
  \BibitemOpen
  \bibfield  {author} {\bibinfo {author} {\bibfnamefont {A.}~\bibnamefont
  {Juodagalvis}}, \bibinfo {author} {\bibfnamefont {J.~M.}\ \bibnamefont
  {Sampaio}}, \bibinfo {author} {\bibfnamefont {K.}~\bibnamefont {Langanke}}, \
  and\ \bibinfo {author} {\bibfnamefont {W.~R.}\ \bibnamefont {Hix}},\ }\href
  {\doibase 10.1088/0954-3899/35/1/014031} {\bibfield  {journal} {\bibinfo
  {journal} {Journal of Physics G: Nuclear and Particle Physics}\ }\textbf
  {\bibinfo {volume} {35}},\ \bibinfo {pages} {014031} (\bibinfo {year}
  {2007})}\BibitemShut {NoStop}%
\bibitem [{\citenamefont {Zhi}\ \emph {et~al.}(2011)\citenamefont {Zhi},
  \citenamefont {Langanke}, \citenamefont {Martínez-Pinedo}, \citenamefont
  {Nowacki},\ and\ \citenamefont {Sieja}}]{Zhi_NPA_2011}%
  \BibitemOpen
  \bibfield  {author} {\bibinfo {author} {\bibfnamefont {Q.}~\bibnamefont
  {Zhi}}, \bibinfo {author} {\bibfnamefont {K.}~\bibnamefont {Langanke}},
  \bibinfo {author} {\bibfnamefont {G.}~\bibnamefont {Martínez-Pinedo}},
  \bibinfo {author} {\bibfnamefont {F.}~\bibnamefont {Nowacki}}, \ and\
  \bibinfo {author} {\bibfnamefont {K.}~\bibnamefont {Sieja}},\ }\href
  {\doibase https://doi.org/10.1016/j.nuclphysa.2011.04.010} {\bibfield
  {journal} {\bibinfo  {journal} {Nuclear Physics A}\ }\textbf {\bibinfo
  {volume} {859}},\ \bibinfo {pages} {172 } (\bibinfo {year}
  {2011})}\BibitemShut {NoStop}%
\bibitem [{\citenamefont {{Lattimer}}\ \emph {et~al.}(1985)\citenamefont
  {{Lattimer}}, \citenamefont {{Burrows}},\ and\ \citenamefont
  {{Yahil}}}]{BurrowsLattimerYahil}%
  \BibitemOpen
  \bibfield  {author} {\bibinfo {author} {\bibfnamefont {J.~M.}\ \bibnamefont
  {{Lattimer}}}, \bibinfo {author} {\bibfnamefont {A.}~\bibnamefont
  {{Burrows}}}, \ and\ \bibinfo {author} {\bibfnamefont {A.}~\bibnamefont
  {{Yahil}}},\ }\href {\doibase 10.1086/162830} {\bibfield  {journal} {\bibinfo
   {journal} {\apj}\ }\textbf {\bibinfo {volume} {288}},\ \bibinfo {pages}
  {644} (\bibinfo {year} {1985})}\BibitemShut {NoStop}%
\bibitem [{\citenamefont {{Burrows}}\ and\ \citenamefont
  {{Lattimer}}(1984)}]{Burrows_84}%
  \BibitemOpen
  \bibfield  {author} {\bibinfo {author} {\bibfnamefont {A.}~\bibnamefont
  {{Burrows}}}\ and\ \bibinfo {author} {\bibfnamefont {J.~M.}\ \bibnamefont
  {{Lattimer}}},\ }\href {\doibase 10.1086/162505} {\bibfield  {journal}
  {\bibinfo  {journal} {\apj}\ }\textbf {\bibinfo {volume} {285}},\ \bibinfo
  {pages} {294} (\bibinfo {year} {1984})}\BibitemShut {NoStop}%
\bibitem [{\citenamefont {Horowitz}(1997)}]{Horowitz_1997}%
  \BibitemOpen
  \bibfield  {author} {\bibinfo {author} {\bibfnamefont {C.~J.}\ \bibnamefont
  {Horowitz}},\ }\href {\doibase 10.1103/PhysRevD.55.4577} {\bibfield
  {journal} {\bibinfo  {journal} {Phys. Rev.}\ }\textbf {\bibinfo {volume}
  {D55}},\ \bibinfo {pages} {4577} (\bibinfo {year} {1997})}\BibitemShut
  {NoStop}%
\bibitem [{\citenamefont {Bruenn}\ and\ \citenamefont
  {Mezzacappa}(1997)}]{Bruenn_1997}%
  \BibitemOpen
  \bibfield  {author} {\bibinfo {author} {\bibfnamefont {S.~W.}\ \bibnamefont
  {Bruenn}}\ and\ \bibinfo {author} {\bibfnamefont {A.}~\bibnamefont
  {Mezzacappa}},\ }\href {\doibase 10.1103/PhysRevD.56.7529} {\bibfield
  {journal} {\bibinfo  {journal} {Phys. Rev.}\ }\textbf {\bibinfo {volume}
  {D56}},\ \bibinfo {pages} {7529} (\bibinfo {year} {1997})}\BibitemShut
  {NoStop}%
\bibitem [{\citenamefont {Sumiyoshi}\ \emph {et~al.}(2005)\citenamefont
  {Sumiyoshi}, \citenamefont {Yamada}, \citenamefont {Suzuki}, \citenamefont
  {Shen}, \citenamefont {Chiba},\ and\ \citenamefont
  {Toki}}]{Sumiyoshi_ApJ_2005}%
  \BibitemOpen
  \bibfield  {author} {\bibinfo {author} {\bibfnamefont {K.}~\bibnamefont
  {Sumiyoshi}}, \bibinfo {author} {\bibfnamefont {S.}~\bibnamefont {Yamada}},
  \bibinfo {author} {\bibfnamefont {H.}~\bibnamefont {Suzuki}}, \bibinfo
  {author} {\bibfnamefont {H.}~\bibnamefont {Shen}}, \bibinfo {author}
  {\bibfnamefont {S.}~\bibnamefont {Chiba}}, \ and\ \bibinfo {author}
  {\bibfnamefont {H.}~\bibnamefont {Toki}},\ }\href {\doibase 10.1086/431788}
  {\bibfield  {journal} {\bibinfo  {journal} {The Astrophysical Journal}\
  }\textbf {\bibinfo {volume} {629}},\ \bibinfo {pages} {922} (\bibinfo {year}
  {2005})}\BibitemShut {NoStop}%
\bibitem [{\citenamefont {Fischer}\ \emph {et~al.}(2014)\citenamefont
  {Fischer}, \citenamefont {Hempel}, \citenamefont {Sagert}, \citenamefont
  {Suwa},\ and\ \citenamefont {Schaffner-Bielich}}]{Fischer_EPJA_2013}%
  \BibitemOpen
  \bibfield  {author} {\bibinfo {author} {\bibfnamefont {T.}~\bibnamefont
  {Fischer}}, \bibinfo {author} {\bibfnamefont {M.}~\bibnamefont {Hempel}},
  \bibinfo {author} {\bibfnamefont {I.}~\bibnamefont {Sagert}}, \bibinfo
  {author} {\bibfnamefont {Y.}~\bibnamefont {Suwa}}, \ and\ \bibinfo {author}
  {\bibfnamefont {J.}~\bibnamefont {Schaffner-Bielich}},\ }\href {\doibase
  10.1140/epja/i2014-14046-5} {\bibfield  {journal} {\bibinfo  {journal} {Eur.
  Phys. J.}\ }\textbf {\bibinfo {volume} {A50}},\ \bibinfo {pages} {46}
  (\bibinfo {year} {2014})},\ \Eprint {http://arxiv.org/abs/1307.6190}
  {arXiv:1307.6190 [astro-ph.HE]} \BibitemShut {NoStop}%
\bibitem [{\citenamefont {Pearson}\ \emph {et~al.}(1996)\citenamefont
  {Pearson}, \citenamefont {Nayak},\ and\ \citenamefont {Goriely}}]{pearson96}%
  \BibitemOpen
  \bibfield  {author} {\bibinfo {author} {\bibfnamefont {J.~M.}\ \bibnamefont
  {Pearson}}, \bibinfo {author} {\bibfnamefont {R.~C.}\ \bibnamefont {Nayak}},
  \ and\ \bibinfo {author} {\bibfnamefont {S.}~\bibnamefont {Goriely}},\ }\href
  {\doibase 10.1016/0370-2693(96)01071-4} {\bibfield  {journal} {\bibinfo
  {journal} {Phys. Lett.}\ }\textbf {\bibinfo {volume} {B387}},\ \bibinfo
  {pages} {455} (\bibinfo {year} {1996})}\BibitemShut {NoStop}%
\end{thebibliography}%

\end{document}